\documentclass[journal]{IEEEtran}
\ifCLASSINFOpdf
\else
\fi

\usepackage{ifpdf}
\usepackage[pdftex]{graphicx}
\usepackage[cmex10]{amsmath}
\usepackage{subfig}
\usepackage{amssymb}
\usepackage{mathrsfs}
\usepackage{amsmath}
\usepackage{indentfirst}
\usepackage{booktabs}
\usepackage{multirow}
\usepackage{graphicx}
\usepackage{epstopdf}
\usepackage{fixltx2e}
\usepackage{tcolorbox}
\usepackage{soul}
\usepackage{units}
\usepackage{mathtools}
\usepackage{caption}
\usepackage{epsfig,graphicx,amssymb,amsmath}
\usepackage{subfig}
\usepackage{diagbox}
\usepackage{color}
\usepackage{setspace}

\usepackage{amsthm}
\usepackage{algorithm}
\usepackage[noend]{algpseudocode}
\usepackage{booktabs}
\usepackage{diagbox}
\usepackage{float}
\usepackage{epstopdf}
\usepackage{ragged2e}
\usepackage{stfloats}
\renewcommand{\raggedright}{\leftskip=0pt \rightskip=0pt plus 0cm}

\ifCLASSINFOpdf
\else

\fi

\usepackage{xcolor}

\usepackage{xpatch}

\begin{document}
%
\title{Multi-AUV Cooperative Underwater Multi-Target Tracking Based on Dynamic-Switching-enabled Multi-Agent Reinforcement Learning}

\author{Shengbo Wang,
Chuan Lin,~\IEEEmembership{Member,~IEEE},
Guangjie Han,~\IEEEmembership{Fellow,~IEEE},
Shengchao Zhu,
Zhixian Li,
Zhenyu Wang,
Yunpeng Ma

\thanks{\emph{Corresponding author: Chuan Lin \& Guangjie Han}} 
\thanks{Shengbo Wang, Zhixian Li and Zhenyu Wang are with Software College, Northeastern University, Shenyang, China (e-mails: puppytagge@gmail.com; liaozhaixian@gmail.com; 15063371639@163.com).}
\thanks{Chuan Lin is with Software College, Northeastern University, Shenyang, China and is also with Key Laboratory of Data Analytics and Optimization for Smart Industry (Northeastern University), Ministry of Education, China (e-mails: chuanlin1988@gmail.com).}
\thanks{Guangjie Han and Shengchao Zhu are with the Department of Internet of Things Engineering, Hohai University, Changzhou, 213022, China. (e-mails:  hanguangjie@gmail.com; zhushengchao77@gmail.com).}
\thanks{Yunpeng Ma is with Jiangsu Key Laboratory of Power Transmission and Distribution Equipment Technology, Hohai University, China (e-mail: yunpengma\_hhu@163.com).}
 	
}

\markboth{Journal of \LaTeX\ Class Files,~Vol.~XX, No.~X, XXX~XXXX}%
{Shell \MakeLowercase{\textit{et al.}}: Bare Demo of IEEEtran.cls for IEEE Journals}
%

\IEEEtitleabstractindextext{%
	\begin{abstract}
In recent years, autonomous underwater vehicle (AUV) swarms are gradually becoming popular and have been widely promoted in ocean exploration or underwater tracking, etc. 
In this paper, we propose a multi-AUV cooperative underwater multi-target tracking algorithm especially when the real underwater factors are taken into account.
We first give normally modelling approach for the underwater sonar-based detection and the ocean current interference on the target tracking process.
Then, based on software-defined networking (SDN), we regard the AUV swarm as a underwater ad-hoc network and propose a hierarchical software-defined multi-AUV reinforcement learning (HSARL) architecture.
Based on the proposed HSARL architecture, we propose the "Dynamic-Switching" mechanism, it includes "Dynamic-Switching Attention" and "Dynamic-Switching Resampling" mechanisms which accelerate the HSARL algorithm's convergence speed and effectively prevents it from getting stuck in a local optimum state.
Additionally, we introduce the reward reshaping mechanism for further accelerating the convergence speed of the proposed HSARL algorithm in early phase.
Finally, based on a proposed AUV classification method, we propose a cooperative tracking algorithm called \textbf{D}ynamic-\textbf{S}witching-\textbf{B}ased \textbf{M}ARL (DSBM)-driven tracking algorithm.
Evaluation results demonstrate that our proposed DSBM tracking algorithm can perform precise underwater multi-target tracking, comparing with many of recent research products in terms of various important metrics.

\end{abstract}
	
	\begin{IEEEkeywords}
		AUV swarm, multi-target tracking, Dynamic-Switching Attention, Dynamic-Switching Resampling, software-defined networking.
    \end{IEEEkeywords}}

\maketitle
\IEEEdisplaynontitleabstractindextext
\IEEEpeerreviewmaketitle

\IEEEpeerreviewmaketitle

\section{Introduction}

\label{sec:introduction}

\IEEEPARstart {C}overing more than 70\% of Earth's surface, the ocean is a vital life-support system with abundant biological and mineral resources \cite{9633006,10244971}. 
With the advancement of technology and our increasing knowledge of the oceans, the need for efficient and intelligent exploration technologies, particularly in ocean environment surveillance \cite{9707741,159741}, disaster prevention, and rescue operations \cite{9618770,112178}, becomes increasingly pressing.
In these years, the rapid development of underwater robots, underwater communication, underwater sensing technologies, has given birth to the Autonomous Underwater Vehicle (AUV).
As of today, AUV has become indispensable tools in a wide range of marine applications especially in underwater target tracking \cite{Wang2023}, due to its ability to autonomously navigate underwater environments.

With the continuous improvement and development of swarm intelligence theory, the control technology of AUV swarms has been greatly propelled and enhanced \cite{9485055,9508160}.
The AUV swarms have demonstrated unique advantages in executing tasks in complex marine environments, e.g., the AUV-swarm-based cooperative underwater data collection \cite{10329958}, cooperative underwater target tracking \cite{9530372}.
Through cooperation manner, the AUVs in AUV swarm can cover broader areas, enhancing the efficiency and accuracy of underwater operation. 
Particularly in underwater cooperative multi-target tracking, the AUV swarms have been demonstrated exceptional capabilities in cooperatively tracking various underwater targets such as underwater biological communities \cite{10214407} or environmental pollutants \cite{9390453}.
This is crucial not only for underwater scientific research such as ecological \cite{10161282} surveys but also for various categories of civil \cite{9681175} and military applications \cite{9532493}, such as supporting maritime search, rescue operations and maritime boundary surveillance, etc.
Notably, recently, the multi-agent intelligence or Multi-Agent Reinforcement Learning (MARL) has been emergent and provides an efficient platform for scheduling the AUV swarm to perform cooperative multi-target tracking.

However, effectively managing and cooperating such a complex multi-agent system, especially in dynamic and unpredictable marine environments, has to face many challenges.
Particularly in underwater multi-target tracking, these challenges mainly manifest in several aspects: 1) \textbf{complex marine environment}: the complexity and dynamism of the marine environment, such as ocean currents interference \cite{9722969,9625977}, this demands AUV swarms to possess high adaptability and flexibility to cope with sudden environmental changes and unknown underwater obstacles; 2) \textbf{limited underwater communication}: the high latency and low bandwidth characteristics of underwater communication, especially underwater acoustic-based communication, limit the efficiency of information exchange among the AUVs in the swarm. 
These limitations prevent the support of flexible swarm cooperation \cite{9356608,9451536,9332276}. 
Thus, it requires a scalable network architecture to re-organize and re-define the information exchange architecture for AUV swarm system; 3) \textbf{heterogeneous AUVs}: due to productivity constraints, AUV swarms often comprise various heterogeneous AUVs, making AUV swarm-based cooperative underwater operation more challenging and impeding support for cooperative swarm intelligence \cite{9498989}; 4) \textbf{scalable deployment}: traditional MARL algorithms suffer from long runtime, slow convergence speeds, and high resource consumption, making MARL challenging to deploy in large-scale AUV swarm.

Additionally, in academia, there are two major challenges in MARL: the high variance problem of traditional strategy estimation methods and the slow convergence problem caused by the high dimensionality of Critic networks in multi-agent systems.

Therefore, how to optimize these algorithms to address these challenges without sacrificing efficiency, has become a focal point of current research.

As a cutting-edge networking approach, Software-Defined Networking (SDN) separates the control plane from the data plane in traditional network architectures \cite{10004961,9343333,9750063,9205627}.
By this architecture, the network control functions are centralized into one or more controllers. 
This allows the network resources together with the network information to be flexibly managed and utilized.
Motivated by this, in this work, we display how to utilize SDN technology to re-define the AUV swarm as a AUV swarm network with self-learning ability.
For dedicated MARL architecture towards AUV swarm network, we propose a MARL algorithm incorporating a Dynamic-Switching mechanism, which demonstrates faster convergence and better performance in multi-target tracking tasks under ocean current interference. 
Totally, this paper mainly makes the following contributions:

\begin{enumerate}[]
	\item We regard the AUV swarm as a AUV swarm network, utilize SDN to optimize AUV swarm network and propose a hierarchical software-defined multi-AUV reinforcement learning architecture;
	\item Based on the proposed MARL architecture, we incorporate Dynamic-Switching mechanism (including both ”Dynamic-Switching Attention” and ”Dynamic-Switching Resampling”) to enhance AUV swarm network's self-learning efficiency and accuracy;
	\item We introduces an AUV formation classification algorithm based on fuzzy logic and rule-based expert systems, to enable the AUV swarm network to perform precise tracking of multiple targets in the interference of ocean currents.
\end{enumerate}

The rest of this paper is organized as follows.
The related works on the paper's theme is surveyed in Sec. \ref{Section:2}; 
In Sec. \ref{Section:3}, the preliminary materials are showcased; 
The proposed HSARL is detailed in Sec. \ref{Section:4}; 
The proposed cooperative tracking algorithm is presented in Sec. \ref{Section:5}; 
In Sec. \ref{Section:6}, various categories of evaluation results are presented; 
Sec. \ref{Section:7} concludes the paper and discusses some future research directions derived from this paper.

\section{Related Works}\label{Section:2}
In this section, we investigate recent advances related to the main research topics, specifically categorized into the following three areas: 1) SDN-based underwater multi-agent system; 2) Single AUV-based target tracking; 3) AUV swarm-based target tracking; 4) AI/ML in underwater multi-agent distributed strategies.

\subsection{SDN-based underwater multi-agent system}\label{Section:2-1}
SDN technology, characterized by its network architecture that separates the control plane from the data forwarding plane, provides strong support for centralized control and flexible management of underwater multi-agent systems. This technology is currently widely applied in fields such as underwater multi-agent networks, opening new possibilities for underwater operations and research.

For instance, in \cite{9634122}, the authors propose the DRSIR model, a deep reinforcement learning approach for routing selection in SDN. This method utilizes path state indicators and demonstrates higher efficiency and intelligence compared to traditional Dijkstra and RSIR algorithms. It can dynamically adjust routing strategies based on changes in network traffic, showcasing practical feasibility and superior performance in SDN routing.

In \cite{9076113}, an SDN-based architecture for AUV underwater wireless networks is introduced to support multi-AUV cooperative search. This architecture achieves network information synchronization, node localization, multi-AUV cooperative control, and intelligent data transmission scheduling through software-defined beaconing, hierarchical localization, cooperative control, and software-defined hybrid data transmission frameworks. Building upon these studies, we leverage SDN technology to enhance the underwater vehicle swarm network system.

\subsection{Single AUV-based target tracking}\label{Section:2-2}
In single AUV-based target tracking, the focus is on enhancing the perception \cite{HUY2023113202}, decision-making \cite{CHEN2021109073}, and control capabilities \cite{FANG2022110452} of an individual AUV to autonomously perform identification and tracking of specific targets. This approach necessitates highly integrated sensor systems and advanced data processing algorithms to adapt to complex underwater environments and achieve high-precision tracking.

In \cite{FANG2022110452}, the authors develop a DRL-based control strategy for X-rudder AUVs, utilizing the DDPG algorithm to enable precise posture control and efficient target tracking. This approach achieves precise three-degree-of-freedom posture control and rapid DRL algorithm deployment, and significantly enhances AUV maneuverability for dynamic underwater target tracking.

In \cite{CHEN2021109073}, the authors introduce an optimized AUV path planning method (PPM-BBD) employing L-SHADE, focusing on energy-efficient diving through decision-making under motion constraints. This approach minimizes energy consumption by up to 9\% compared to conventional methods, demonstrating the importance of strategic decision-making in solitary AUV target tracking amidst environmental challenges.

\subsection{AUV swarm-based target tracking}\label{Section:2-3}
AUV swarm-based target tracking leverages the coordinated efforts of multiple AUVs, utilizing principles of swarm intelligence to improve tracking range, efficiency, and robustness. This method not only extends beyond the limitations of a single AUV but also emphasizes distributed decision-making and information sharing among the AUVs to effectively track multiple or wide-area targets in dynamic environments \cite{9685323,110495}.

In \cite{9685323}, the authors propose a novel multi-AUV cooperative tracking scheme named CTDE, which employs a central training and distributed execution approach to achieve safe and effective tracking of moving targets using MADDPG. By conducting central training within a designed secure private network, information sharing is not required during task execution, thereby enhancing the security of the entire system.

In \cite{110495}, the authors discuss trajectory tracking control of multiple AUVs using discrete-time control methods in weak communication environments. The study focuses on considering communication delays (bounded and unbounded) and packet loss scenarios, ensuring stable tracking of given trajectories by designing methods based on leader-follower and virtual leader approaches.

\subsection{AI/ML in underwater multi-agent distributed strategies}\label{Section:2-4}
Normally, a distributed multi-agent strategy for AUV-based underwater systems involves the coordination of multiple Autonomous Underwater Vehicles (AUVs) to accomplish complex tasks. This approach leverages distributed computing and communication networks, enabling each AUV to share environmental information and decisions in real-time, thereby enhancing the efficiency and accuracy of task execution.

In \cite{9206115}, the authors discuss various AI/ML techniques and their specific applications in 6G, including Supervised Learning, Unsupervised Learning, Deep Learning (DL), and Reinforcement Learning (RL). 
These methods also find applications in distributed multi-agent strategies for underwater systems, especially Unsupervised Learning and RL.

\textbf{Unsupervised Learning:} In \cite{10008249}, the authors employ the K-Means clustering algorithm for network partitioning and swarm head selection based on network topology and residual energy of nodes. This aims to achieve efficient data aggregation in underwater sensor networks, thereby improving data transmission efficiency and network energy utilization.

\textbf{Reinforcement Learning (RL):} In \cite{10329958}, the authors propose a target uncertainty map assisted data collection scheme for AUV swarms based on the multi-agent proximal policy optimization (MAPPO) algorithm. 
This approach utilizes current and past search and collection results to establish a target uncertainty map. 
Combined with artificial potential fields, it ultimately significantly enhances data collection efficiency and accuracy.

Totally, multi-AUV-based underwater target tracking technology has received widespread research attention. 
However, this research field still faces several challenges, such as adaptability to dynamic and complex marine environments, flexibility and robustness of AUV swarm networks, and how to stably and accurately track multiple targets. 
In response to these issues, our work employ the SDN technology to improve the scalability and functionality of the AUV swarm system.
On account of the above system architecture, we propose a Dynamic-Switching-enabled MARL architecture which is more feasible for AUV swarm system's self-learning.
And, we propose as well a novel AUV classification method to improve the scalability of the AUV swarm-based tracking system.

\section{Preliminary Materials}\label{Section:3}
In order to accurately track underwater targets and adapt to the dynamic factors in real underwater environments, we consider the effects of sonar detection and ocean current interference.
Furthermore, we transform the target tracking scenario based on AUV swarm network into a Markov Decision Process (MDP).

\subsection{Sonar detection modeling}\label{Section:3-1}

In real complex underwater environments, traditional electromagnetic-based detection methods are limited in their effectiveness. Therefore, in this paper, we utilize sonar technology to detect AUVs' position together with their tracking targets. 
Equipped with sonar, AUVs emit directional sound waves to scan their surrounding environment. 
This process involves using a sector array sonar to capture echo signals from different directions, and the variations in signal intensity allow us to estimate the position of targets.  
In this study, a specific active sonar equation model is employed to simulate and optimize the target detection process, which can be expressed by Eq. \ref{EQ1}.
\begin{equation}
\label{EQ1}
EM = SL - 2TL + TS - (NL - DI) - DT
\end{equation} 
where \( SL \) is the strength of the sonar signal emitted, \( TL \) is the attenuation of sound propagation, \( TS \) is the target echo strength, \( NL \) is the ambient marine noise, \( DI \) is the system's noise suppression capability, and \( DT \) is the minimum signal-to-noise ratio required for the equipment to function properly. 
The units for the notations mentioned are all in decibels (dB).

\subsection{Ocean current modeling}\label{Section:3-2}

The underwater environments' uncertainty, especially the influence of ocean currents, poses significant challenges to the navigation and positioning capabilities of AUVs. 
It demands AUVs to possess a high level of adaptability and intelligence.

The Navier-Stokes equations are the fundamental equations for describing fluid dynamics behavior. 
They can simulate the characteristics of fluid flow.
Additionally, they provide a computational environment for calculating the forces exerted by ocean currents (Eq. \ref{EQ2}).

\begin{equation}
\label{EQ2}
\rho \left( \frac{\partial \mathbf{u}}{\partial t} + \mathbf{u} \cdot \nabla \mathbf{u} \right) = -\nabla p + \mu \nabla^2 \mathbf{u} + \mathbf{F}
\end{equation}
where \( \rho \) is fluid density, \( \mathbf{u} \) is velocity, \( \frac{\partial \mathbf{u}}{\partial t} \) is velocity change over time, \( \mathbf{u} \cdot \nabla \mathbf{u} \) is self-advection, \( \nabla p \) is pressure change, \( \mu \) is viscosity, \( \nabla^2 \mathbf{u} \) is the diffusion term, and \( \mathbf{F} \) is external forces.

However, the Navier-Stokes equations involve a large number of partial differential calculations, resulting in high computational costs and long processing times. 
To reduce the complexity and better handle ocean current, we use the Reynolds-Averaged Navier-Stokes (RANS) equations. 
The RANS equations are derived from the Navier-Stokes equations by applying a time or statistical averaging process. 
This averaging effectively reduces computational complexity and has gained widespread use in industrial applications.
The RANS equations decompose the velocity into mean and fluctuating components: \(u = \overline{u} + u'\), where \(\overline{u}\) is the mean velocity and \(u'\) is the fluctuating velocity. The RANS equations can be expressed as follows:

\begin{equation}
\label{EQ33}
\rho \left( \frac{\partial \overline{u}}{\partial t} + \overline{u} \cdot \nabla \overline{u} \right) = -\nabla \overline{p} + \mu \nabla^2 \overline{u} + \nabla \cdot \left( -\rho \overline{u' u'} \right) + F
\end{equation}

After calculating \(\overline{u}\) and \(\frac{\partial \overline{u}}{\partial t}\) through the RANS equations, we use the drag equation (Eq. \ref{EQ4}), lift equation (Eq. \ref{EQ5}), and virtual mass equation (Eq. \ref{EQ6}) to calculate the forces acting on the structure under the influence of ocean currents:

\begin{equation}
\label{EQ4}
F_D = \frac{1}{2}\rho u^2 C_D A
\end{equation}
where \( F_D \) is the drag force, \( \rho \) is fluid density, \( u \) is relative fluid velocity, \( C_D \) is drag coefficient, and \( A \) is object's frontal area.

\begin{equation}
\label{EQ5}
F_L = \frac{1}{2}\rho u^2 C_L A
\end{equation}
where \( F_L \) is the lift force, \( C_L \) is lift coefficient.

\begin{equation}
\label{EQ6}
F_{VM} = \rho C_{VM} V \frac{\partial \mathbf{u}}{\partial t}
\end{equation}
where \( F_{VM} \) is the virtual mass force, \( C_{VM} \) is the virtual mass coefficient, \( V \) is AUV's volume, and \( \frac{\partial u}{\partial t} \) is the rate of change of velocity.

\begin{figure}
    \centering
    \includegraphics[width=0.5\linewidth]{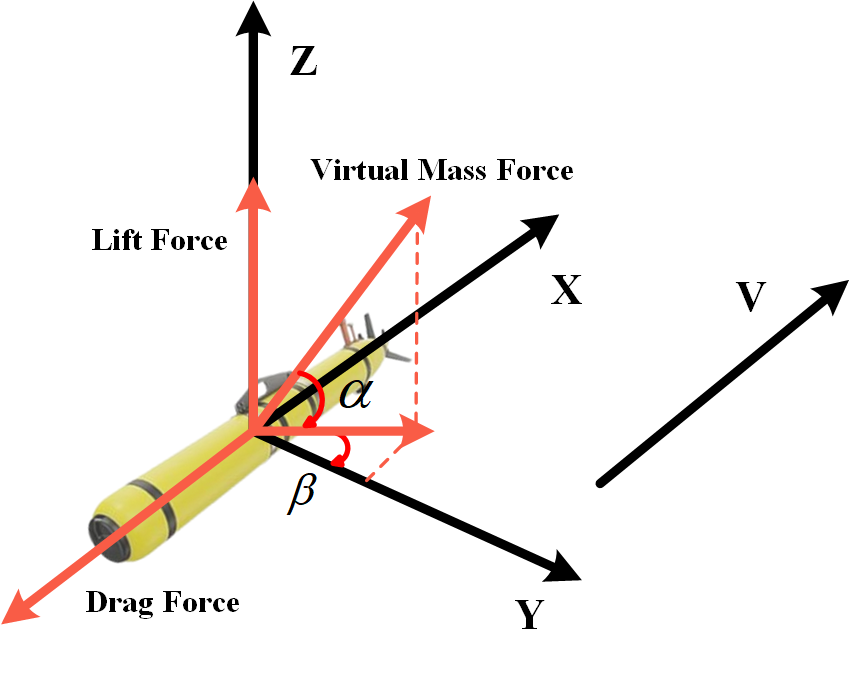}
    \caption{Schematic diagram of ocean current}
    \label{fig11}
\end{figure}

Through the above equations (Eq. \ref{EQ4}-Eq. \ref{EQ6}), we can compute the complex and variable forces exerted by ocean currents on the AUV. 
The three forces are shown in Fig. \ref{fig11}. 
The overall calculation for the ocean current can be summarized in Algorithm 1.

\begin{algorithm}
\small
\caption{RANS-based ocean current interference calculation algorithm}
\begin{algorithmic}[1]
\State Initialize the Set \( N_E \), \( T \) respectively.
\State Initialize AUV's frontal area \( A \), volume V, drag coefficient \( C_D \), lift coefficient \( C_L \), and inertia coefficient \( C_{VM} \).

\For{\( \text{Episode} = 1 \) to \( N_E \)}
    \State Initialize fluid density \( \rho \), dynamic viscosity \( \mu \), and external forces \( \mathbf{F} \).
    \State Initialize environment and fluid velocity \( \mathbf{u} \) and pressure gradient \( \nabla p \).
    \For{\( t = 1 \) to \( T \)}
        \State Calculate the velocity time derivative \( \frac{\partial \mathbf{u}}{\partial t} \) by RANS equation.
        \State Update fluid velocity \( \mathbf{u} \) and pressure gradient \( \nabla p \).
        \State Calculate the drag force \( F_D \), lift force \( F_L \), and virtual mass force \( F_{VM} \) by Eq. \ref{EQ4} to Eq. \ref{EQ6}.
        \State Combine forces to determine net force on the AUV.
    \EndFor
\EndFor

\end{algorithmic}
\end{algorithm}

\subsection{Markov decision process modeling}\label{Section:3-3}

In the decision-making process of AUV swarm network, the environment is defined as a MDP (Eq. \ref{EQ7}), consisting of five main components: the state set \(\mathcal{S}\), the action set \(\mathcal{A}\), the state transition probabilities \(\mathcal{P}\), the reward function \(\mathcal{R}\), and the discount factor \(\gamma\):

\begin{equation}
\label{EQ7}
\mathcal{M} = (\mathcal{S}, \mathcal{A}, \mathcal{P}, \mathcal{R}, \gamma)
\end{equation}

The state set \(\mathcal{S}\) encompasses the sonar detection results in different coordinate systems and their location information. Specifically, it is a set composed of the target and the AUV's echo intensity along with their location information. \(\mathcal{S}\) can be represented as \(\mathcal{S} = \{s_1, \ldots, s_N\}\), where \(s_i = (\eta_i, \phi_i, o_i)\), \(o_i = (\kappa_i, \sigma_i) \in \mathbb{R}^{N_\kappa + N_\sigma}\), \(\kappa_i\) represents the echo intensity from the target, \(\sigma_i\) represents the echo intensity from other AUVs, \(N_\kappa\) is the number of targets being tracked, and \(N_\sigma\) is the number of other AUVs.
The action set \(\mathcal{A}\) in this study represents the discrete action set of all AUVs, denoted as \(\mathcal{A}=\{a_1,\ldots,a_N\}\), where each action \(a_i\) is a vector representing seven possible actions (e.g., up, down, left, right, forward, backward, and stay in place).
The state transition probabilities \(\mathcal{P}\) describe the dynamic transition relationship between states, which is key for the algorithm to understand how different states transition into each other.
The discount factor \(\gamma\), where the value of \(\gamma\) ranges between 0 and 1, together determines the calculation method for long-term benefits in policy evaluation.
Further, the reward function \(\mathcal{R}\) will be explained in Sec. \ref{Section:5}, where we consider tracking accuracy, collision avoidance, ocean current stability, and energy efficiency to enhance multi-AUV cooperative tracking effects.

\section{Hierarchical Software-Defined Multi-AUV Reinforcement Learning Architecture}\label{Section:4}

In this section, we introduce \textbf{H}ierarchical \textbf{S}oftware-defined multi-\textbf{A}UV \textbf{R}einforcement \textbf{L}earning (HSARL), which is a software-defined MARL architecture based on centralized training and decentralized execution.

 \begin{figure*}[bth]
 	\centering
 	\includegraphics[width=0.7\linewidth]{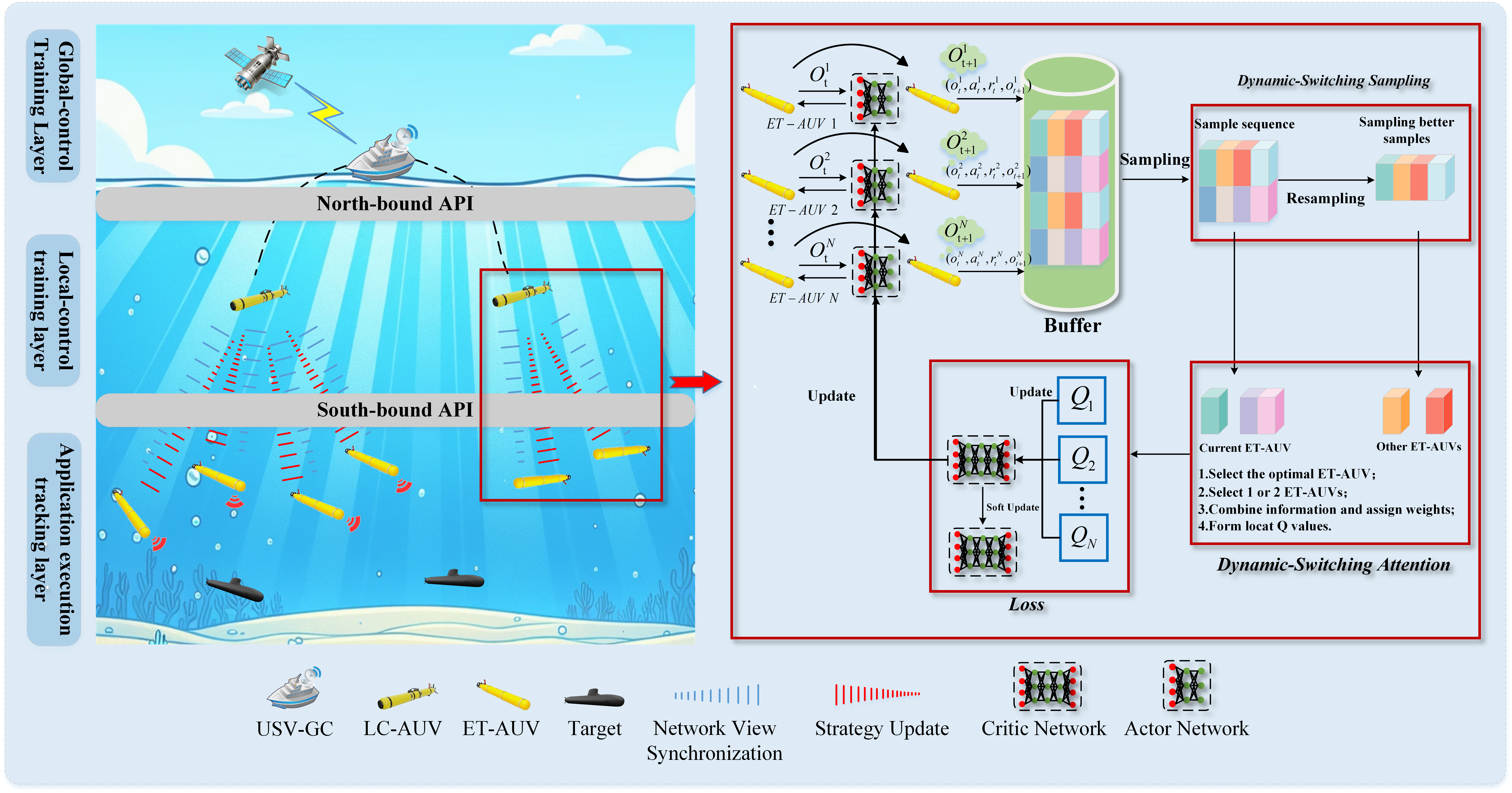}
	\caption{Proposed HSARL}
 	\label{fig111}
\end{figure*}

\subsection{Overview of HSARL}\label{Section:4-1}

HSARL framework consists of three functional layers: global-control training layer, local-control training layer, and application execution tracking layer, which can be detailed as the following.

\textbf{Global-control training layer:} The global-control training layer comprises a global controller based on Unmanned Surface Vessel-based Global Controller (USV-GC). 
It aims to coordinate Local Controller-based AUV (LC-AUV)s in various regions through a dedicated northbound interface, ensuring efficient classification of large-scale tasks to suitable local-control training layers by deconstructing and optimizing tasks. 
The strategy updates and task scheduling mechanism in this layer not only support the efficient execution of complex tasks but also continuously optimize the global network view by periodically obtaining status information from the lower layers.
This ensures the flexibility and efficiency of network operations, especially when facing environmental challenges and changing task requirements.

\textbf{Local-control training layer:} It focuses on optimizing tasks issued by the global-control training layer into specific strategies through training, and conveying them to each managed Execution Tracking-based AUV (ET-AUV). 
Upon assignment of subregion tasks by the global-control training layer, each LC-AUV begins centralized local training. 
This process uses the unique characteristics of the ET-AUV subset it manages, following the parameters and tasks provided by the global-control training layer.
This further refines tasks and formulates specific execution strategies to guide its managed ET-AUVs. 
Additionally, it periodically maintains the local network view through a dedicated southbound interface. 
As LC-AUVs require a certain amount of computation, they are typically equipped with high-performance batteries and computing units. 
This training mode ensures that all ET-AUVs can execute tasks according to uniform standards and optimal strategies, while also allowing for rapid adjustments tailored to local environments and specific task requirements.

\textbf{Application execution tracking layer:} It is directly responsible for executing specific strategies. 
This layer consists of ordinary ET-AUVs swarms.
Each ET-AUV independently executes tasks and strategies received from the local-control training layer. 
Therefore, communication among ET-AUVs is not imposed. 
This idea (the idea of "separate control from operation" in SDN) reduces communication costs and enhances the flexibility and robustness of the AUV swarm network. 
Even if individual ET-AUVs fails or are destroyed, it does not affect the overall efficiency of the network.

Overall, HSARL reduces the coupling between individual AUVs, addressing the scalability issues present in traditional architectures.
The HSARL framework is summarized in Fig. \ref{fig111}.
The entire task classification processing is summarized in Fig. \ref{fig1}. 

\subsection{Beacon framework for HSARL}\label{Section:4-2}

In order to ensure precise target tracking based on the proposed HSARL, it is crucial to establish both local and global network views for LC-AUVs and USV-GC. 
Therefore, we propose an SDN beacon framework to synchronize and share information among ET-AUVs, LC-AUVs, and USV-GC. 
The proposed SDN beacon framework can be carried out by the following:

\textbf{Phase 1}: To update local and global network views, the beaconing is activated between USV-GC/LC-AUVs and their subordinate levels within the region requesting synchronization (SYN) information. 
Upon receiving a synchronization request, each LC-AUV/ET-AUV replies with key information such as its AUV ID and status to update the network views.

\textbf{Phase 2}: To allocate specific-tasks/strategies to LC-AUVs/ET-AUVs through centralized training. 
In each training round, USV-GC/LC-AUVs send requests to LC-AUVs/ET-AUVs they manage after training completion, and send specific-tasks/strategies to them through operation requests.
Each LC-AUV/ET-AUV then replies with key information.

In summary, Phase 1 describes the network view updating procedure, while Phase 2 displays the task classification processing. 
To improve the information exchange efficiency, the reply information in Phase 2 also includes information for updating the network views.  
Therefore, these replies are also considered as an update to the network views.

\section{Proposed Cooperative Tracking Algorithm}\label{Section:5}
In this section, we showcase our proposed cooperative tracking algorithm.
Especially, we firstly propose a classification algorithm called AUV Scalable Management and Allocation (ASMA) to classify/allocate ET-AUVs to each formation for performing cooperative multi-target tracking. 
To enhance the tracking accuracy and the convergence speed of the AUV swarm network, on account of HSARL, we employ the concept of "Dynamic-Switching", integrate the "Dynamic-Switching Attention" mechanism into the Actor-Critic model and develop the \textbf{D}ymiac-\textbf{S}witching-\textbf{B}ased \textbf{M}ARL (DSBM)-driven tracking algorithm.
Additionally, to improve the efficiency of sample utilization in DSBM, we introduce a novel "Dynamic-Switching Resampling" mechanism based on experience replay buffer mechanism.

\subsection{AUV classification method}\label{Section:5-1}
In underwater multi-target tracking problems, efficiently and reasonably classifying tracking formations (consisting of a series of ET-AUVs) for each underwater target is an important issue. 
Totally, the proposed ASMA achieves efficient and accurate formation assignments for ET-AUVs by utilizing a combination of fuzzy logic and rule-based expert systems, integrating various performance indicators.

The evaluation process initially employs fuzzy logic scoring to map specific performance indicators of ET-AUVs onto a membership degree. 
The mathematical expression for this function is as follows (Eq. \ref{EQ8}):

\begin{figure}[bth]
	\centering
	\includegraphics[width=0.7\linewidth]{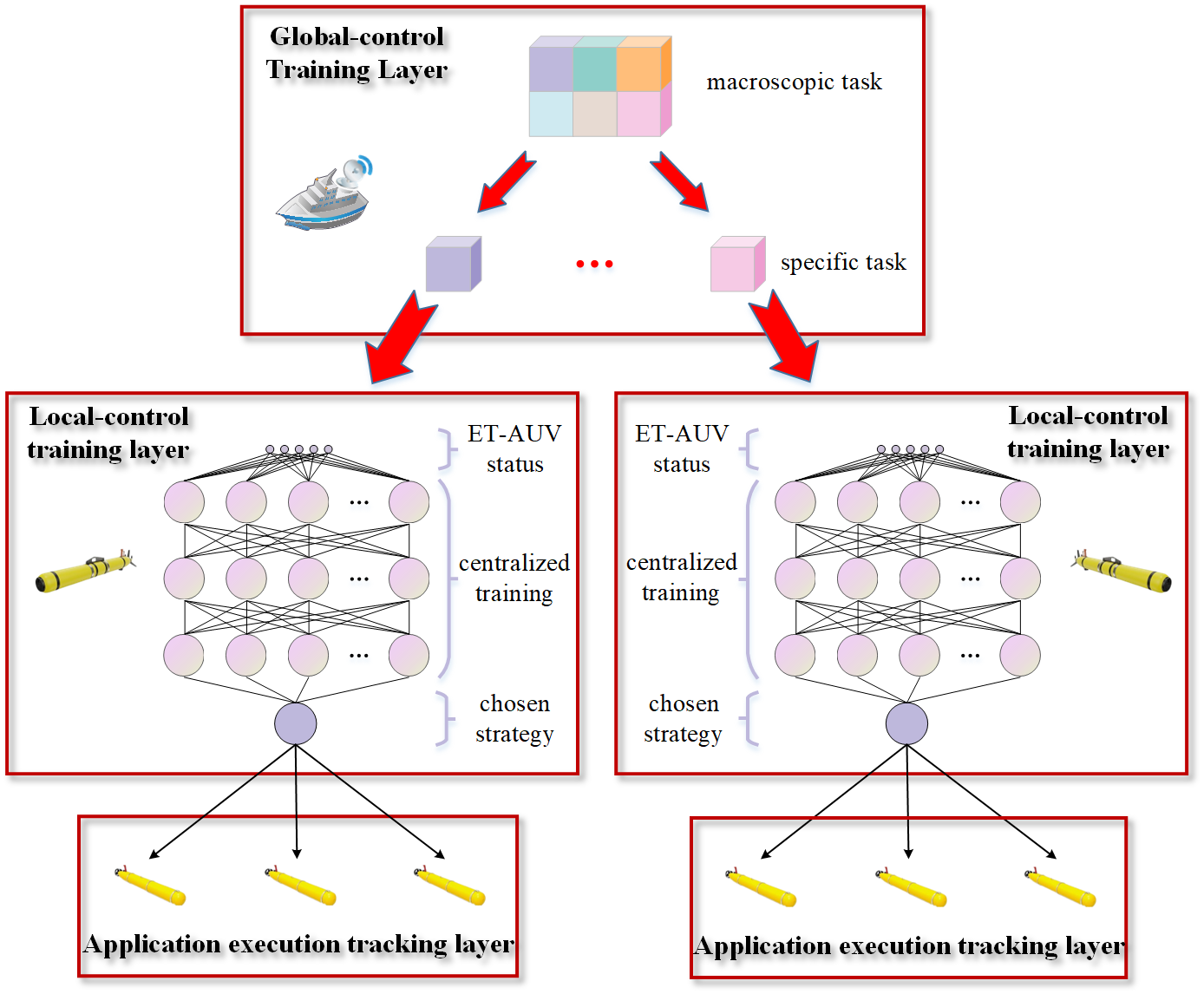}
	\caption{Task classification based on HSARL}
	\label{fig1}
\end{figure}

\begin{table}[H]
\small
\centering
\caption{Special notations in Sec. \ref{Section:5}}
\label{table1}
\begin{tabular}{cc}
\hline
\textbf{Notation} & \textbf{Description} \\ \hline
\( \alpha_i \) & Acceleration score \\
\( \beta_i \) & Remaining battery score \\
\( \kappa_i \) & Carrying capacity score \\
\( \epsilon_i \) & Energy consumption score \\
\( d_{i,j} \) & Distance between ET-AUV \(i\) to the target \( j \)\\
\( D_i \) & Perception range \\
\( v_i \) & Speed score \\ \hline
\end{tabular}
\end{table}

\begin{equation}
\label{EQ8}
\mu(x) = \begin{cases} 
0 & \text{if } x < \text{range\_min} \\
1 & \text{if } x > \text{range\_max} \\
\frac{x - \text{range\_min}}{\text{range\_max} - \text{range\_min}} & \text{otherwise}
\end{cases}
\end{equation}
where \(\mu(x)\) is the membership degree of notation \(x\) normalized between 0 and 1, \(x\) is the actual value of the notation, and \(\text{range\_min}\) and \(\text{range\_max}\) set the normalization range for \(x\), specifying the minimum and maximum values, respectively.

The rule-based expert system takes the fuzzy membership degrees as input. 
It then calculates the overall score \(\mathit{S}_{i,j}\) acquired by ET-AUV \(i\) and target \(j\).
The notations and their meanings are shown in {Table \ref{table1}. 

Thus, based on ASMA, the score \(\mathit{S}_{i,j}\) acquired by ET-AUV \(i\) and underwater tracking target \(j\) can be computed by the following steps:

\noindent\textbf{Step 1: Energy regulation score:}

\begin{equation}
\label{EQ9}
\mathit{S}_{i,j} += \left( v_i^3 + 2 \alpha_i^2 \right) \times \frac{1}{1 + e^{-\beta_i}} \times \left( 1 - \frac{d_{i,j}}{D_i} \right)
\end{equation}

Significant increases in total \(\mathit{S}_{i,j}\) occur when ET-AUVs have fast speeds, stable accelerations, and sufficient power. 
However, when the remaining power is low, even if the ET-AUV accelerates quickly, the \(\mathit{S}_{i,j}\) will be lower to avoid that ET-AUVs quickly deplete their power.
Additionally, as the distance between the ET-AUV and the target approaches the maximum perception range, the score decreases to prevent losing track of the target.

\noindent\textbf{Step 2: Efficiency load balancing score:}

\begin{equation}
\label{EQ10}
\mathit{S}_{i,j} += \cosh(\epsilon_i - \kappa_i) \times \sqrt{\lvert \kappa_i - \epsilon_i \rvert}
\end{equation}

By comparing energy consumption with payload capacity, the balance between energy efficiency and stability of ET-AUVs during task execution is emphasized.

\noindent\textbf{Step 3: Comprehensive performance score:}

\begin{equation}
\label{EQ11}
\mathit{S}_{i,j} += \frac{v_i^2 + \alpha_i^2 + \epsilon_i^2}{3}
\end{equation}

This rule considers the comprehensive impact of speed, acceleration, and energy consumption, reflecting the stability of ET-AUV's overall performance under different operating conditions.

\noindent\textbf{Step 4: Dynamic counter-reaction score:}

\begin{equation}
\label{EQ12}
\mathit{S}_{i,j} += \frac{1}{1 + e^{\alpha_i - v_i}}
\end{equation}

If the ET-AUV can smoothly adjust its speed, it indicates high-performance control capability under dynamic conditions. 
This is crucial for EC-AUVs to perform complex underwater evasive action, e.g., collision avoidance, precise positioning, efficient navigation, etc.

\noindent\textbf{Step 5: Fine operational expertise score:}

\begin{equation} 
\label{EQ13}
\mathit{S}_{i,j} += \text{if } v \leq 0.3 \text{ then } v \times \kappa_i \text{ else } 0
\end{equation}

In underwater tasks requiring precise control, ET-AUVs with low speed and high payload capacity will receive additional \(\mathit{S}_{i,j}\) due to higher operational accuracy.

\noindent\textbf{Step 6: Remote rapid response score:}

\begin{equation}
\label{EQ14}
\mathit{S}_{i,j} += \text{if } v \geq 0.7 \text{ then } v \times \beta_i \times \frac{d_{i,j}}{D_i} \text{ else } 0
\end{equation}

In underwater urgent tasks requiring the ET-AUV to fleetly arrive at the distant targets, high-speed ET-AUVs at greater distances from the target receive additional \(\mathit{S}_{i,j}\), reflecting their rapid response capability to distant targets.

Totally, the proposed ASMA is carried out based on the \(\mathit{S}_{i,j}\) of each ET-AUV for each target. 
This ensures that every ET-AUV tracks a target and each target is tracked by at least one ET-AUV. 
The optimization goal is to maximize the cumulative score.

In summary, ASMA can be summarized as Algorithm 2, which considers various special factors in different scenarios, classifying rational and efficient tracking formations.

\begin{algorithm}[htb]
\small
    \caption{ET-AUV performance classification algorithm}
    \label{alg:1}
        \begin{algorithmic}[1]
        \State Initialize the number \( k \) of ET-AUVs and the number \( T \) of targets.
        \For{\( i = 1 \) to \( k \)}
            \State Initialize ET-AUV \( i \)'s \( \alpha_i \), \( \beta_i \), \( \kappa_i \), \( \epsilon_i \), \( d_{i,j} \), \( D_i \) and \( v_i \).
        \EndFor
        \For{\( i = 1 \) to \( k \)}
            \State Calculate membership values for ET-AUV \( i \) by Eq. \ref{EQ8}.
        \EndFor
        \For{\( i = 1 \) to \( k \)}
            \For{\( j = 1 \) to \( T \)}
                \State Update the score \( \mathit{S}_{i,j} \) of ET-AUV \( i \) for target \( j \) by Eq. \ref{EQ9} to Eq. \ref{EQ14}.
            \EndFor
        \EndFor
        \State Assign the best formation plan based on score \( \mathit{S}_{i,j} \).
    \end{algorithmic}
\end{algorithm}

\subsection{Proposed cooperative tracking policy}\label{Section:5-2}
In this section, we introduce the proposed reward function under the HSARL reinforcement learning architecture. 
In reinforcement learning, by modeling the problem as MDP, the problem can be transformed into a corresponding maximization of the cumulative expected reward. 
Therefore, we define the reward function as follows (Eq. \ref{EQ24}):

\begin{equation}
\label{EQ24}
R_i(s, a) = \alpha r_i^{\delta} + \beta r_i^{\zeta} + \delta r_i^{\mu} + \gamma r_i^{\sigma}
\end{equation}
where  \( r_i^{\delta} \) is the tracking reward, \( r_i^{\zeta} \) is the collision-avoidance reward, \( r_i^{\mu} \) is the ocean current stability reward, and \( r_i^{\sigma} \) is the velocity matching reward.

For ET-AUV \( i \), the tracking reward \( r_i^{\delta} \) is designed to control the distance between EC-AUV and the tracking target, thereby maintaining the ET-AUV and the target within the best tracking distance range (Eq. \ref{EQ25}).

\begin{equation}
\label{EQ25}
r_i^{\delta} = -\omega_1 \frac{|d^\phi_{(i,j)} - d^\phi_{\text{be}}|}{d^\phi_{\text{be}}}
\end{equation}
where \( d^\phi_{(i,j)} \) is the distance between ET-AUV \( i \) and target \( j \), and \( d^\phi_{\text{be}} \) is the best tracking distance.

At the same time, to ensure a safe distance between ET-AUVs during the tracking process, we introduce a collision-avoidance reward \( r_i^{\zeta} \) (Eq. \ref{EQ26}):

\begin{equation}
\label{EQ26}
r_i^{\zeta} = -\omega_2 \sum_{\substack{j=1 \\ j \neq i}}^{N} \frac{|d^{\psi}_{(i,j)} - d^{\psi}_{\text{be}}|}{d^{\psi}_{\text{be}}}
\end{equation}
where \( d^{\psi}_{(i,j)} \) is the distance between ET-AUV \( i \) and \( j \), and \( d^{\psi}_{\text{be}} \) is the best distance between ET-AUVs.

To promote the smooth movement of ET-AUVs under the influence of ocean currents, we introduce an ocean current stability reward \( r_i^{\mu} \) (Eq. \ref{EQ27}). 
This reward helps mitigate the effects on ET-AUV's velocity caused by ocean currents.

\begin{equation}
\label{EQ27}
r_i^{\mu} = -\omega_3 \frac{\lVert v_{\text{cur}}(i) - v_{\text{pre}}(i) \rVert}{\lVert v_{\text{pre}}(i) \rVert}
\end{equation}
where \( v_{\text{cur}}(i) \) is ET-AUV \( i \)'s current velocity, \( v_{\text{pre}}(i) \) is ET-AUV \( i \)'s previous velocity.

To avoid excessive energy consumption by ET-AUVs due to frequent acceleration and deceleration, we introduce a velocity matching reward \( r_i^{\sigma} \) (Eq. \ref{EQ28}):

\begin{equation}
\label{EQ28}
r_i^{\sigma} = -\omega_4 \frac{\lVert v_{\text{age}}(i) - v_{\text{tar}}(j) \rVert}{\lVert v_{\text{tar}}(j) \rVert}
\end{equation}
where \( v_{\text{age}}(i) \) is the velocity of the ET-AUV, \( v_{\text{tar}}(j) \) is the velocity of the target.

\subsection{Proposed DSBM-driven tracking algorithm}\label{Section:5-3}
\subsubsection{Proposed DSBM}
Normally, as it is aforementioned in Sec. \ref{sec:introduction}, there are two major challenges: the high variance problem of traditional strategy estimation methods and the slow convergence problem caused by the high dimensionality of Critic networks in multi-agent systems. 
To address these issues, in this paper, we introduce temporal difference techniques to reduce evaluation variance and propose DSBM, which adopts the "Dynamic-Switching" mechanism to improve learning efficiency and accelerate policy convergence. 
The "Dynamic-Switching" mechanism focuses on optimizing the handling of high-value samples, as shown on the right side of Fig. \ref{fig111}.

\textbf{Dynamic-Switching Attention mechanism:} During the process of tracking targets, there are a total of \( N \) ET-AUVs. 
For a specific ET-AUV \( i \), we need to calculate its Q-value function \( Q_{\psi_i}(o, a) \), where \( o \) represents the current state and \( a \) represents the action. 
The \( Q_{\psi_i}(o, a) \), which represents the value of the action taken by ET-AUV \( i \) given a particular observation and action.
It can be obtained through the following steps.

For each ET-AUV, first identify the ET-AUV with the highest reward in the current training round (\( AUVbe \)) (Eq. \ref{EQ15}).
\begin{equation}
\label{EQ15}
AUV_{\text{be}} = \text{argmax} (R_{f})
\end{equation}
where \( R_{f} \) is the set of reward value for the AUVs at round \( f \).

After that, it randomly select the experience from the remaining ET-AUVs (Eq. \ref{EQ16}).
\begin{equation}
\label{EQ16}
AUV_{\text{ad}} = 
\begin{cases} 
AUV_{\text{be}} \cup \text{ran}(R^*, 2) & \text{if } i = AUV_{\text{be}} \\
AUV_{\text{be}} \cup \text{ran}(R^*, 1) & \text{if } i \neq AUV_{\text{be}}
\end{cases}
\end{equation}
where \( AUV_{\text{ad}} \) is the chosen ET-AUVs for update, and \( \text{ran}(X, n) \) selects \( n \) experience from \( X \) randomly. 
The set \( R^* \) denotes all ET-AUVs excluding \( AUV_{\text{be}} \).

The specific selection process is as follows:

\begin{enumerate}[]
\item If the current ET-AUV has the highest reward, then randomly select two from the remaining ET-AUVs to avoid the algorithm getting stuck in local optimum state.

\item If it doesn't, then only randomly select one from the other ET-AUVs to increase the proportion of experiences from the best ET-AUV.
\end{enumerate}

By the following Proof A, it can be seen that the aforementioned part of the proposed Dynamic-Switching Attention mechanism is able to lead to better training results by leveraging high-reward samples for more effective policy updates.

After obtaining \( AUVad \), exclude \( AUVad \) from the ET-AUV set to obtain \( AUVord \) (Eq. \ref{EQ17}).

\begin{equation}
\label{EQ17}
AUV_{\text{ord}} = R_f \backslash AUV_{\text{ad}}
\end{equation}

Then, concatenate the state and action information of ET-AUVs in \( AUVad \) to form \( m1 \) (Eq. \ref{EQ18}).

\begin{equation}
\label{EQ18}
    m_{1}=cat_{k\in AUV_{ad}}(o_{k},a_{k})
\end{equation}

Apply neural networks and Dynamic-Switching Attention weight processing to ET-AUVs in \( AUVord \) to form \( m2 \) (Eq. \ref{EQ19}).

\begin{equation}
\label{EQ19}
m_2 = \sum_{k \in AUV_{ord}} \omega_{k} f(o_k, a_k)
\end{equation}
where the Dynamic-Switching Attention weight \( \omega_{k} \), is to evaluate the importance of the information of the other AUVs, and \( f(o_k, a_k) \) is a function to extract key features from \( o_k \) and \( a_k \) of the \( k \)-th ET-AUV. 

\begin{tcolorbox} 
\footnotesize
\textit{Proof A}
\normalsize
\tcblower 

\footnotesize
To demonstrate that the dynamic-switching attention mechanism accelerates convergence speed, we consider the standard Q-value update for ET-AUV \( i \):

\begin{equation}
\label{EQ45}
Q_{t+1}^i(s, a) = Q_t^i(s, a) + \alpha \delta_t^i
\end{equation}

where the temporal difference (TD) error is defined as:

\begin{equation}
\label{EQ46}
\delta_t^i = r_t^i + \gamma \max_{a'} Q_t^i(s', a') - Q_t^i(s, a)
\end{equation}

With the attention mechanism, the expected TD error becomes a weighted sum of the errors from the best ET-AUV and the others:

\begin{equation}
\label{EQ47}
\mathbb{E}[\delta_t^i] = p_{\text{be}} \mathbb{E}[\delta^{\text{be}}] + (1 - p_{\text{be}}) \mathbb{E}[\delta^{\text{others}}]
\end{equation}

where \( p_{\text{be}} \) is the probability of selecting experiences from the best ET-AUV. Since the best ET-AUV has superior performance:

\begin{equation}
\label{EQ48}
\mathbb{E}[\delta^{\text{be}}] < \mathbb{E}[\delta^{\text{others}}]
\end{equation}

Therefore, increasing \( p_{\text{be}} \) leads to a decrease in the expected TD error:

\begin{equation}
\label{EQ49}
p_{\text{be}} \uparrow \implies \mathbb{E}[\delta_t^i] \downarrow
\end{equation}

A lower expected TD error results in faster convergence of the Q-values:

\begin{equation}
\label{EQ51}
\text{Convergence speed} \propto \frac{1}{\mathbb{E}[\delta_t^i]}
\end{equation}

Thus, the dynamic-switching attention mechanism accelerates the model's convergence by increasing \( p_{\text{be}} \) and reducing \( \mathbb{E}[\delta_t^i] \).

\normalsize 
\end{tcolorbox}

\begin{equation}
\label{EQ50}
    \omega_{k}=\frac{r_{fk}}{\sum_{m\in AUV_{ord}}r_{fm} }
\end{equation}
where \( r_{fk} \) is the reward of \( k \)-th ET-AUV at \( f \)-th round.

Finally, input \( m1 \) and \( m2 \) into a neural network (MLP) to calculate the Q-value function (Eq. \ref{EQ20}).

\begin{equation}
\label{EQ20}
Q^\psi_i(o, a) = f_i(m_1, m_2)
\end{equation}
where \(f_{i}\) is a three-layer MLP. 

Totally, by the following Proof B, it can be concluded that the proposed dynamic weighting phase in Eq. \ref{EQ50} is able to reduce the variance in policy estimation by dynamically assigning higher weights to high-reward experiences.

\textbf{Dynamic-Switching Resampling mechanism:} In our study, to enhance the training efficiency by using better samples, we propose an improved sampling strategy based on the experience replay buffer \( D \). 
The buffer \( D \) accumulates data from multiple training rounds. 
This data is stored in the form of a series of tuples \( D_i \), where each tuple records the states, actions, and rewards of all ET-AUVs in a specific round.

Our approach involves extracting a subset \( D_1 \) comprising \( n \) samples from \( D \) in each training round, which is then used to update the Critic network. 
Subsequently, the update process selects the top 50\% highest-reward samples from \( D_1 \). 
These selected samples form a new set \( D_2 \), whose samples are then used for a second update for the Critic network.

\textbf{Network updates:} Further, we employ a progressive strategy to update network parameters using the soft update method (Eq. \ref{EQ21}), gradually adjusting the target network parameters to stabilize the training process.

\begin{tcolorbox} 
\footnotesize
\textit{Proof B}
\normalsize 
\tcblower 
\footnotesize
Let \( Q(s, a) \) be the state-action value function, and \( Q_{\text{target}}(s, a) \) be the target value for the update. In traditional Q-learning, the variance of the policy estimation is given by:

\begin{equation}
\label{EQ41}
\sigma^2 = \mathbb{E} \left[ \left( Q_{\text{target}}(s, a) - \mathbb{E}[Q_{\text{target}}(s, a)] \right)^2 \right]
\end{equation}

In the Dynamic-Switching Attention mechanism, dynamic weights \( \omega_i \) are introduced, and the update rule becomes:

\begin{equation}
\label{EQ42}
Q(s, a) \leftarrow Q(s, a) + \alpha \left( \sum_{i=1}^{N} \omega_i Q_{\text{target},i}(s, a) - Q(s, a) \right)
\end{equation}

where \( \omega_i = \frac{R_i}{\sum_{j=1}^{N} R_j} \), dynamically adjusted based on the rewards \( R_i \). The new variance \( \sigma^2_{\text{new}} \) is:

\begin{equation}
\label{EQ43}
\sigma^2_{\text{new}} = \sum_{i=1}^{N} \omega_i^2 \sigma^2_i
\end{equation}

Since \( \omega_i \) is larger for higher rewards \( R_i \), the contribution of low-value samples is reduced. This leads to:

\begin{equation}
\label{EQ44}
\sigma^2_{\text{new}} < \sigma^2_{\text{original}}
\end{equation}

Thus, the dynamic weighting reduces the variance in policy estimation.

\normalsize 
\end{tcolorbox}

\begin{equation}
\label{EQ21}
\hat{\omega} \leftarrow \tau \omega + (1 - \tau) \hat{\omega}
\end{equation}
where \(\omega\) represents the parameters of the original network and \(\bar{\omega}\) represents the parameters of the target work.

Since all Critic networks share common features, we establish a shared loss function to centrally optimize all the Critics (Eq. \ref{EQ22}). 
This loss function (regarded as a measure of training effectiveness) is based on the sum of differences between expected and actual Q-values.

\begin{equation}
\label{EQ22}
\begin{array}{c}
\mathcal{L}_{Q}(\psi)=\sum_{i=1}^{N} E_{\left(o, a, r,  o^{\prime}\right)}\left[\left(Q_{i}^{\psi}(o, a)-y_{i}\right)^{2}\right], 
\end{array}
\end{equation}
where \begin{equation}
    y_{i}=r_{i}+\gamma Q_{i}^{\bar{\psi}}\left(\mathrm{o}^{\prime}, a^{\prime}\right)
\end{equation}

To slightly adjust the policy of each ET-AUV, we utilize gradient ascent to update their policy parameters (Eq. \ref{EQ23}).

\begin{equation}
\label{EQ23}
\nabla_{\theta_{i}} J\left(\pi_{\theta}\right)=E_{o, a \sim \mathcal{D}}\left[\nabla_{\theta_{i}} \pi_{\theta_{i}}\left(a_{i} \mid o_{i}\right) \nabla_{a_{i}} Q_{i}^{\psi}(o, a)\right]
\end{equation}
where \( a_i = \pi_{\theta_i}(o_i) \).

\textbf{Reward reshaping mechanism:} 
In actual situations, the tracking reward (Eq. \ref{EQ25}) plays the most significant role in the reward function. 
The shorter the distance between the AUV and the target, the smaller the negative reward. 

However, in the early tracking stage, the distance between the AUV and the target is too large. 
As a result, AUV’s any action has a little effect on the system reward, making it easily falls into a local optimum state.
Therefore, we introduce a reward reshaping mechanism and propose the reward reshaping term \( r_i^{\phi} \) (Eq. \ref{EQ29}):
\begin{equation}
\label{EQ29}
\phi=\frac{\left(p_{\mathrm{tar}}-p_{\mathrm{age}}\right)\cdot v_{\mathrm{age}}}{|p_{\mathrm{tar}}-p_{\mathrm{age}}|\cdot|v_{\mathrm{age}}|}
\end{equation}
where \( p_{\text{tar}} \) is the position vector of the target, \( p_{\text{age}} \) is the position vector of the AUV, and \( v_{\text{age}} \) is the velocity vector of the AUV.

The reward reshaping term is determined by the angle between the AUV's velocity direction and the relative direction to the target. 
The smaller the angle, the higher the reward.

Finally, we introduce the reward reshaping mechanism into the reward function.
We obtain the reshaped reward function \(Re(s, a)\) as follows:

\begin{equation}
\label{EQ30}
Re(s, a) = R_i(s, a) + \phi
\end{equation}

\subsubsection{Proposed cooperative tracking algorithm}
Based on the AUV swarm network, the cooperative underwater target tracking algorithm can be summarized as shown in Algorithm 3. 

The proposed Algorithm 3 combines the proposed DSBM architecture with the proposed tracking policy. 
This integration enables accurate and cooperative performance in underwater cooperative target tracking, while the ET-AUV's endurance and resistance to ocean current interference is taken into account.

\begin{algorithm}
\small
\caption{HSARL-based DSBM-driven tracking algorithm}
\begin{algorithmic}[1]
\State Initialize the Set \( N_E \), \( T \), \( N \), respectively.
\State Initialize the weights of Actor and Critic networks corresponding to each ET-AUV.
\State Initialize a random process \( \mathcal{N} \) for exploration of action.
\State Initialize replay buffer, environment.

\For{\( \text{Episode} = 1 \) to \( N_E \)}
    \State Reset environment.
    \State Assign tracking formations according to Algorithm 2.
    \For{\( t = 1 \) to \( T \)}
        \State Each ET-AUV takes action \( a_i = \pi_{\theta_i}(o_i) + \mathcal{N}_t \) based on the policy and exploration.
        \State Calculate current force according to Algorithm 1.
        \State Update environment, obtain next state and reward.
        \State Add experience to the replay buffer.
        \For{\( i = 1 \) to \( N \)}
            \State Select ET-AUVs based on Dynamic-Switching Attention mechanism.
            \State Calculate \( Q_i(\psi) \) based on selected ET-AUVs.
            \State Update the actor according to Eq. \ref{EQ23}.
            \State Update the Critic according to Eq. \ref{EQ22}.
        \EndFor
    \EndFor
    \State Perform Dynamic-Switching Resampling mechanism from \( (x_t, a_t, r_t, x_{t+1}) \).
    \State Update the target network of each ET-AUV according to Eq. \ref{EQ21}.
\EndFor

\end{algorithmic}
\end{algorithm}

\section{Evaluations}\label{Section:6}
This section presents the evaluation results, where we conduct an in-depth analysis of the newly proposed tracking algorithm from two perspectives: the effectiveness of the algorithm and the performance of target tracking.
We demonstrate our proposed algorithm by comparing it with some of current mainstream MARL algorithms.
\begin{table}[H]
\small
\centering
\caption{Simulation parameters}
\label{table2}
\begin{tabular}{ccc}
\hline
\textbf{Notation} & \textbf{Description} & \textbf{Value} \\ \hline
\( N_A \) & Number of ET-AUVs & [4,6,12] \\
\( N_T \) & Number of Targets & [2,3,4] \\
\( l_r \) & Learning rate & \( 3e-3 \) \\
\( N_E \) & Number of training rounds & 5000 \\
\( N_h \) & Number of neurons in the hidden layer & 64 \\
\( \gamma \) & Discount factor & 0.95 \\
\( \tau \) & Network update coefficient & \( 1e-2 \) \\
\( d^{\phi}_{\text{min}} \) & Minimum tracking distance & 80 m \\
\( d^{\kappa}_{\text{min}} \) & Minimum distance among ET-AUVs & 80 m \\

{\( L_{e} \)} & {Episode length} & {600} \\
{\( B_{s} \)} & {Buffer size} & {100000} \\
{\( U_{i} \)} & {Update interval} & {2000} \\
{\( M_{s} \)} & {Minimal buffer size to update} & {4000} \\
{\( B \)} & {Batch size} & {512} \\
{\( \rho \)} & {Fluid density} & {\( 1000 \, \text{kg/m}^3 \)} \\
{\( \mu \)} & {Fluid viscosity} & {1e-3 Pa·s} \\
    {\( D \)} & {Damping factor} & {0.25} \\
{\( \Delta t \)} & {Simulation time step} & {0.1 s} \\
\hline
\end{tabular}
\end{table}

\subsection{Simulation setup}\label{Section:6-1}
The test is conducted on a computer equipped with an Intel(R) Core(TM) i9-12900H, 2.50GHz processor and a RAM of 32GB.

We use MPE\cite{mordatch2017emergence} as the simulation environment. 
Based on this,we construct a model using coordinate gridization, treating ET-AUVs as moving particles in a 3D underwater environment. 
The targets initially locate near the origin and move with a random velocity, while the ET-AUVs are uniformly distributed within a region one kilometer away from the targets.

We conduct simulations under three scenarios: 4 ET-AUVs tracking 2 targets, 6 ET-AUVs tracking 3 targets, and 12 ET-AUVs tracking 4 targets. 
Each scenario includes the conditions both with and without ocean current interference, respectively.

The relevant evaluation parameters are detailed in {Table \ref{table2}.

\begin{figure*}[bth]
	\centering 
	\subfloat[4 ET-AUVs tracking 2 targets without ocean current interference]
	{\includegraphics[width=0.28\textwidth]{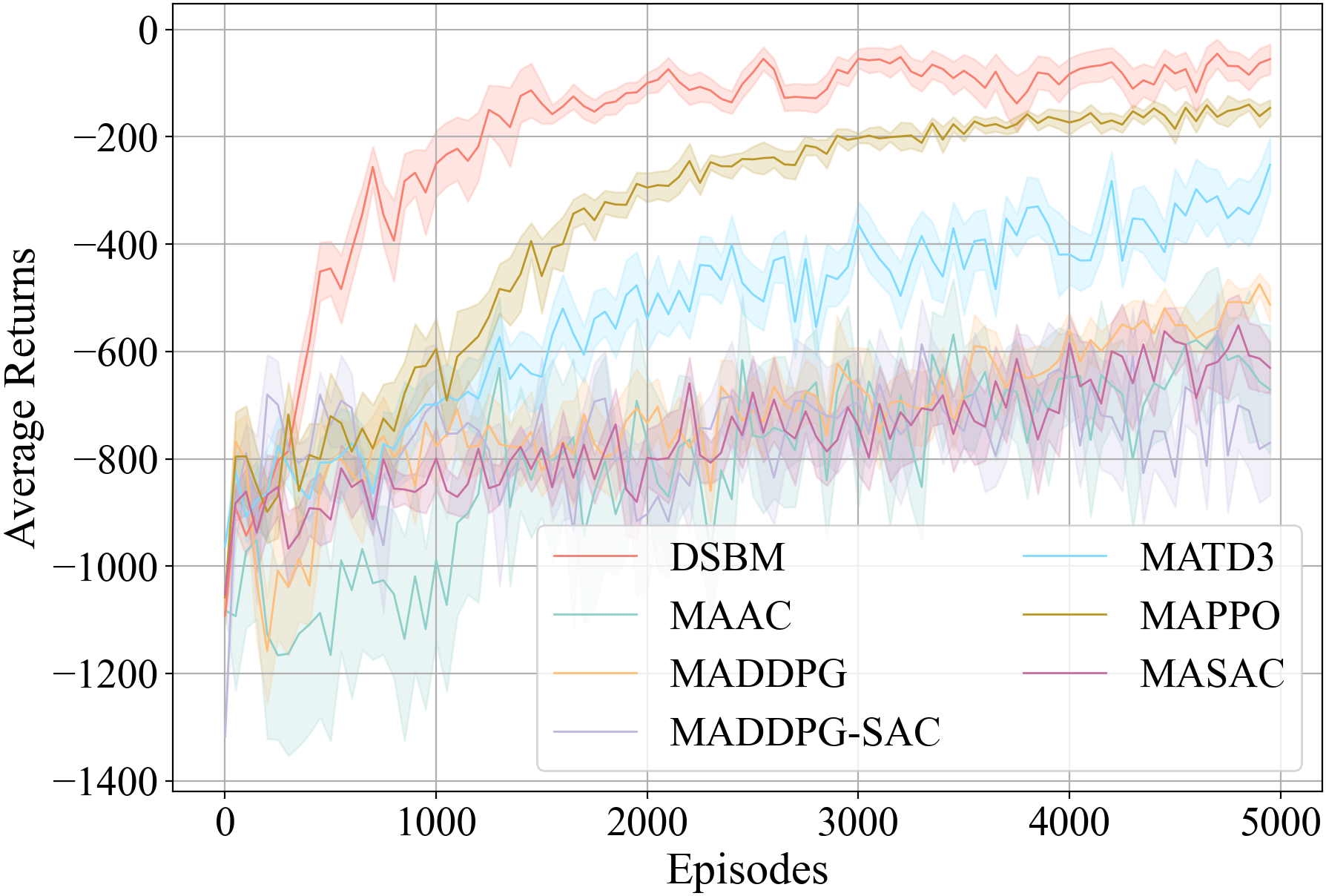}\label{fig2}}
	\hfil
	\subfloat[4 ET-AUVs tracking 2 targets with ocean current interference]
	{\includegraphics[width=0.28\textwidth]{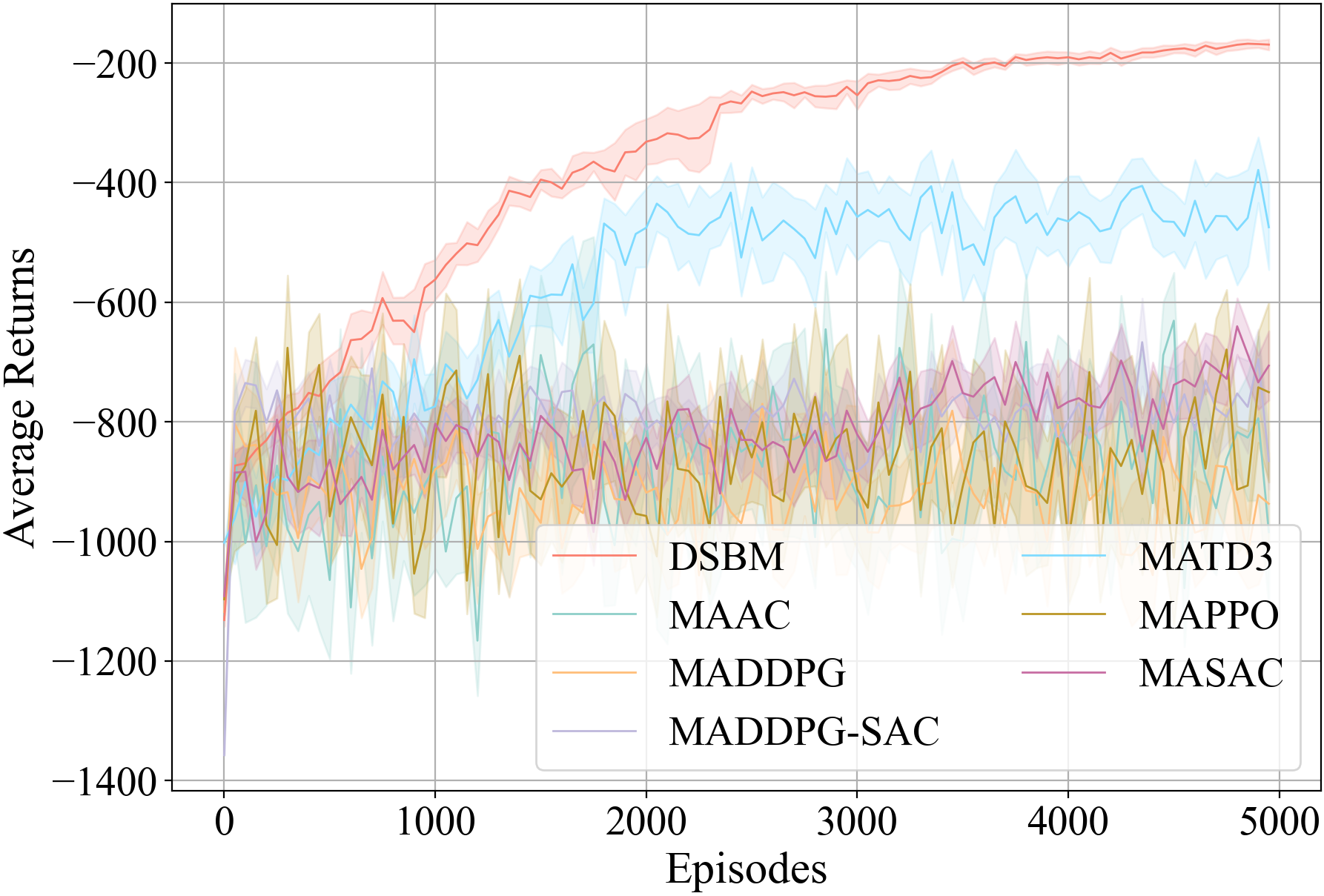}\label{fig3}}
        \hfil
        \subfloat[12 ET-AUVs tracking 4 targets without ocean current interference]
	{\includegraphics[width=0.28\textwidth]{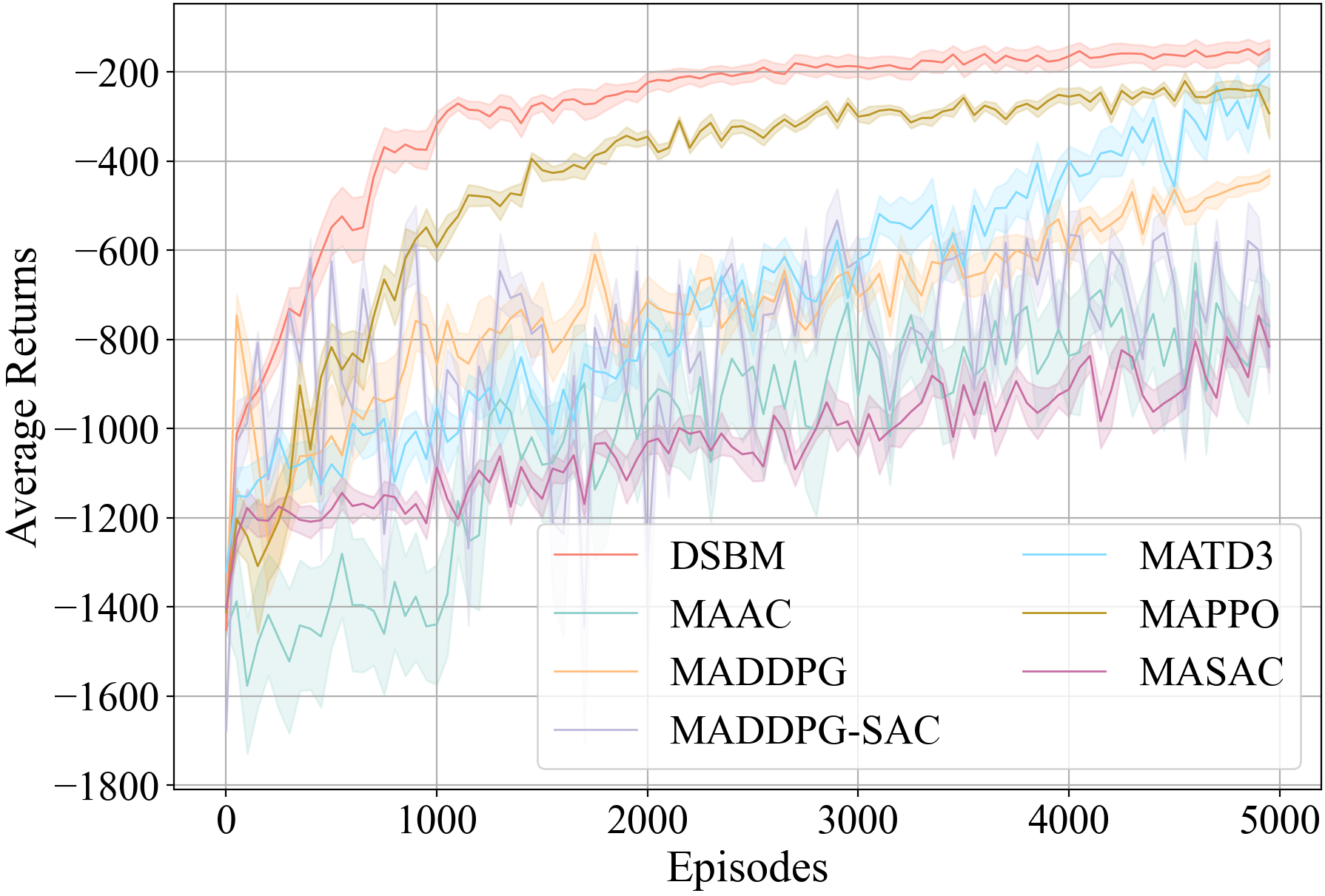}\label{fig12}}
	\hfil
	\subfloat[12 ET-AUVs tracking 4 targets with ocean current interference]
	{\includegraphics[width=0.28\textwidth]{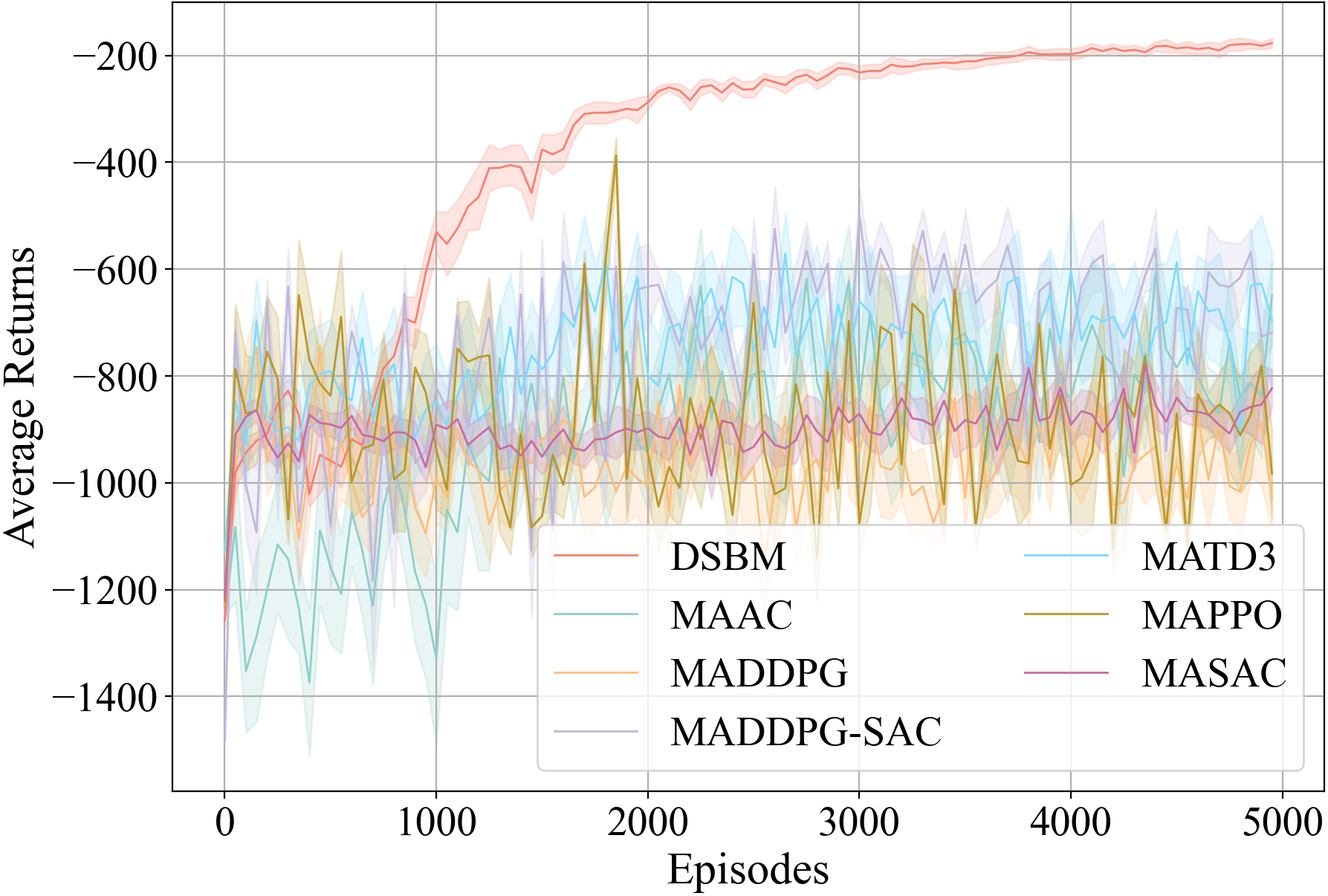}\label{fig13}}
 	\hfil
	\subfloat[4 ET-AUVs tracking 2 targets without ocean current interference]
	{\includegraphics[width=0.28\textwidth]{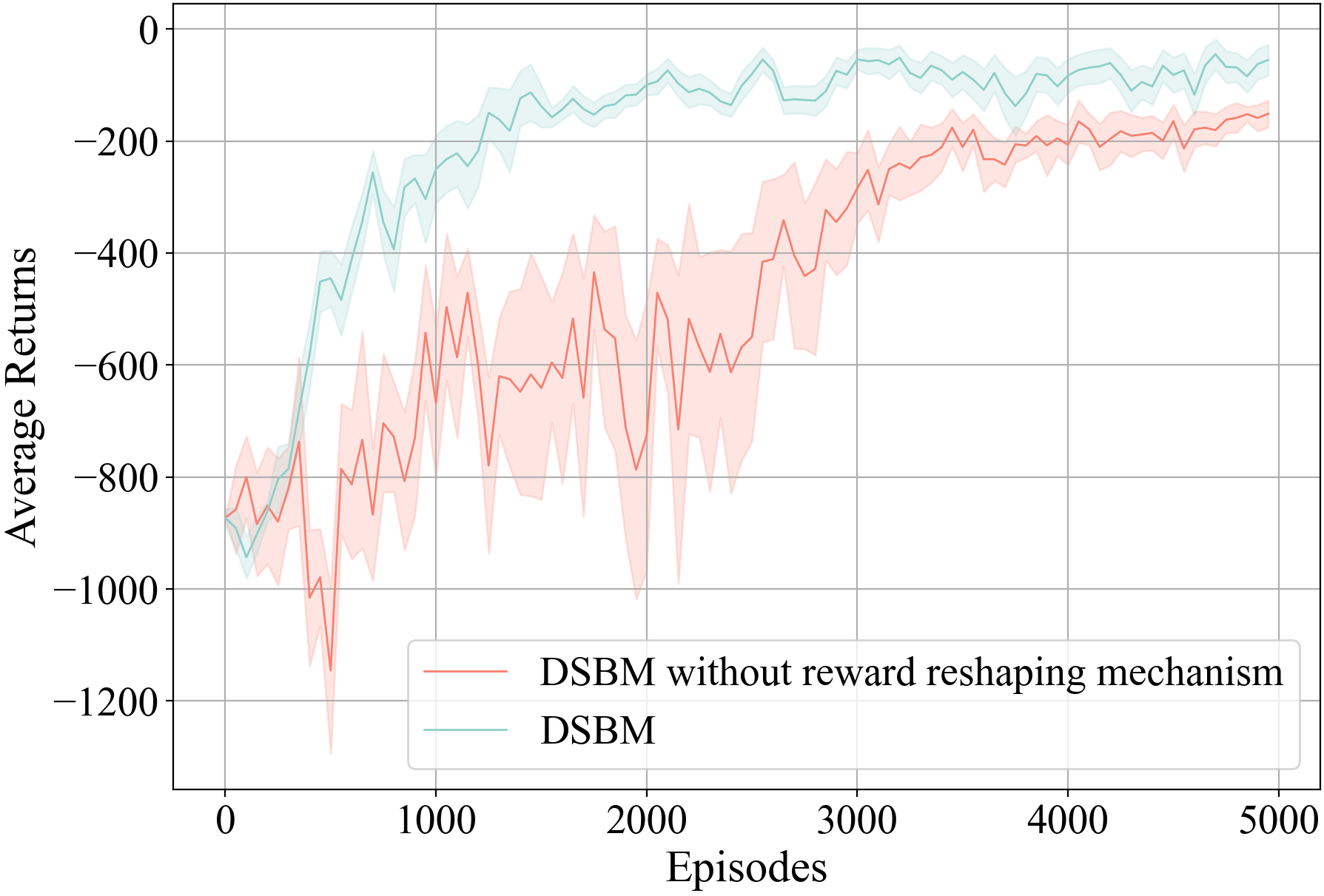}\label{fig-comparison1}}
	\hfil
	\subfloat[12 ET-AUVs tracking 4 targets without ocean current interference]
	{\includegraphics[width=0.28\textwidth]{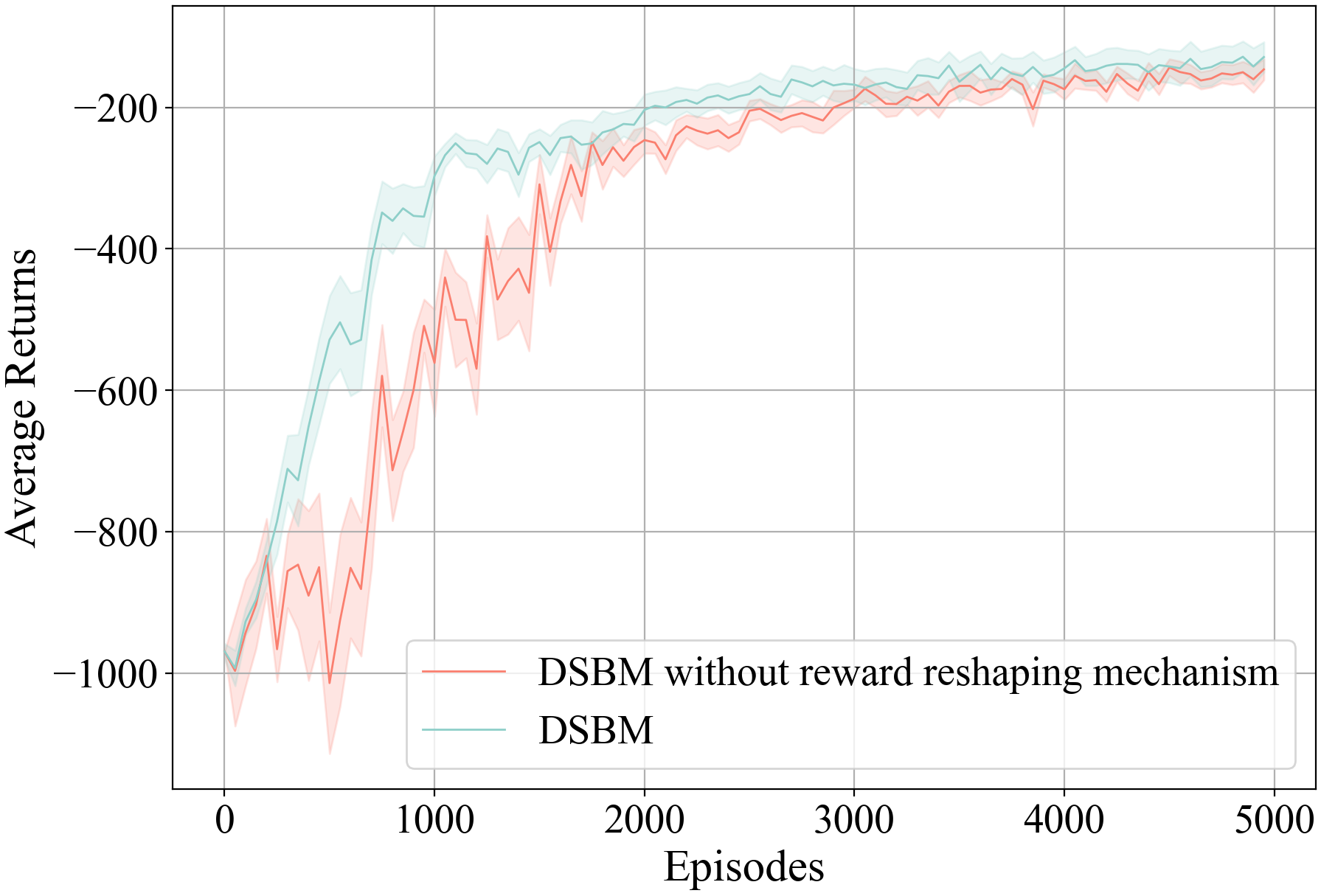}\label{fig-comparison2}}
 \caption{Convergence speed comparison}
\end{figure*}

\subsection{Results}\label{Section:6-2}

First, to measure the effectiveness of the proposed DSBM, we compare DSBM, with MADDPG \cite{37}, MAAC \cite{38}, MADDPG-SAC \cite{39}, MATD3 \cite{ackermann2019reducing}, MAPPO \cite{yu2022the}, and MASAC \cite{9446746}. 
Note that MADDPG-SAC is based on MADDPG where the entropy value (that is adaptively adjusted by SAC), is incorporated into the optimization goal to change the update of the Critic network.
We compare DSBM with MADDPG-SAC to evaluate the significance of entropy value in the algorithm comparison.
We conduct the comparison by respectively utilizing different algorithms to track the underwater targets in terms of convergence speed, tracking distance maintenance, velocity difference maintenance, intra-formation strategy consistency, energy consumption, tracing accuracy, respectively.
 
\subsubsection{Convergence speed}
To evaluate the convergence of the proposed algorithm, we repeat multiple experiments in different tracking scenarios and select samples within a 95\% confidence interval. Overall, as shown in Fig. \ref{fig2}, Fig. \ref{fig3}, Fig. \ref{fig12}, and Fig. \ref{fig13}, DBSM outperforms the other six methods in convergence speed.

Firstly, it should be noted, in Fig. \ref{fig2}, Fig. \ref{fig3}, and Fig. \ref{fig12}, MATD3 clearly performs better than the other four algorithms, except for DBSM and MAPPO. This is due to MATD3’s noise injection mechanism. Adding noise to the policy network’s output (i.e., actions) encourages the agent to explore more in the state-action space. This helps the model avoid falling into local optima prematurely and gather more experience, leading to faster convergence.

Secondly, we note as well our algorithm occasionally falls into local optima but quickly recovered, as shown in the early stages of convergence line in Fig. \ref{fig13}. This is due to that our proposed Dynamic-Switching Attention mechanism, which selects information from the best-performance ET-AUV in the current round, randomly chooses information from one or two other ET-AUVs, and then extracts features from the remaining ET-AUVs. This approach increases the proportion of information from the best-performance ET-AUVs in the training data, leading to better training results in similar scenarios and faster recovery from the early exploration phase. 
For more details about this, please refer to Proof A in Sec. \ref{Section:5-3} of this paper.

Thirdly, our proposed algorithm showcases the fastest training speed during the early stages, as displayed in Fig. \ref{fig2}, Fig. \ref{fig3}, Fig. \ref{fig12}, and Fig. \ref{fig13}. This is due to that our proposed Dynamic-Switching Resampling mechanism, which selects the top 50\% of data based on reward values for updating. This help accelerate finding the optimal solution.

In summary, the results in Fig. \ref{fig2}, Fig. \ref{fig3}, Fig. \ref{fig12}, and Fig. \ref{fig13} demonstrate that DBSM focuses more on the information from well-performing ET-AUVs, leading to more meaningful exploration and significantly reduces the time spent on ineffective exploration.

Additionally, to test the availability of the reward reshaping mechanism in DSBM, we conduct as well a convergence speed comparison between DSBM and DSBM without reward reshaping mechanism.
After introducing the reward reshaping term, DSBM's convergence improves significantly. As shown in Fig. \ref{fig-comparison1} and Fig. \ref{fig-comparison2}, the green curve represents the convergence under the proposed DSBM. It is obvious that the DSBM converges quickly than the DSBM (without reward reshaping mechanism) and is not easy to fall into a local optimum.
That's because, without reward reshaping mechanism, the tracking reward (in Eq.\ref{EQ25}) plays the most significant role in the reward function. The shorter the distance between the AUV and the target, the smaller the negative reward. However, in the early tracking stage, the distance between the AUV and the target is too large. As a result, AUV’s any action has a little effect on the system reward, making it easily falls into a local optimum state.
After incorporating the reward reshaping mechanism, it encourages ET-AUV to move toward the target, especially in the early training stages. This effectively shortens the initial exploration phase, speeds up convergence, and prevents getting stuck in local optimum state.

\begin{figure}[bth]
	\centering 
	\subfloat[12 ET-AUVs tracking 4 targets without ocean current interference]
	{\includegraphics[width=0.24\textwidth]{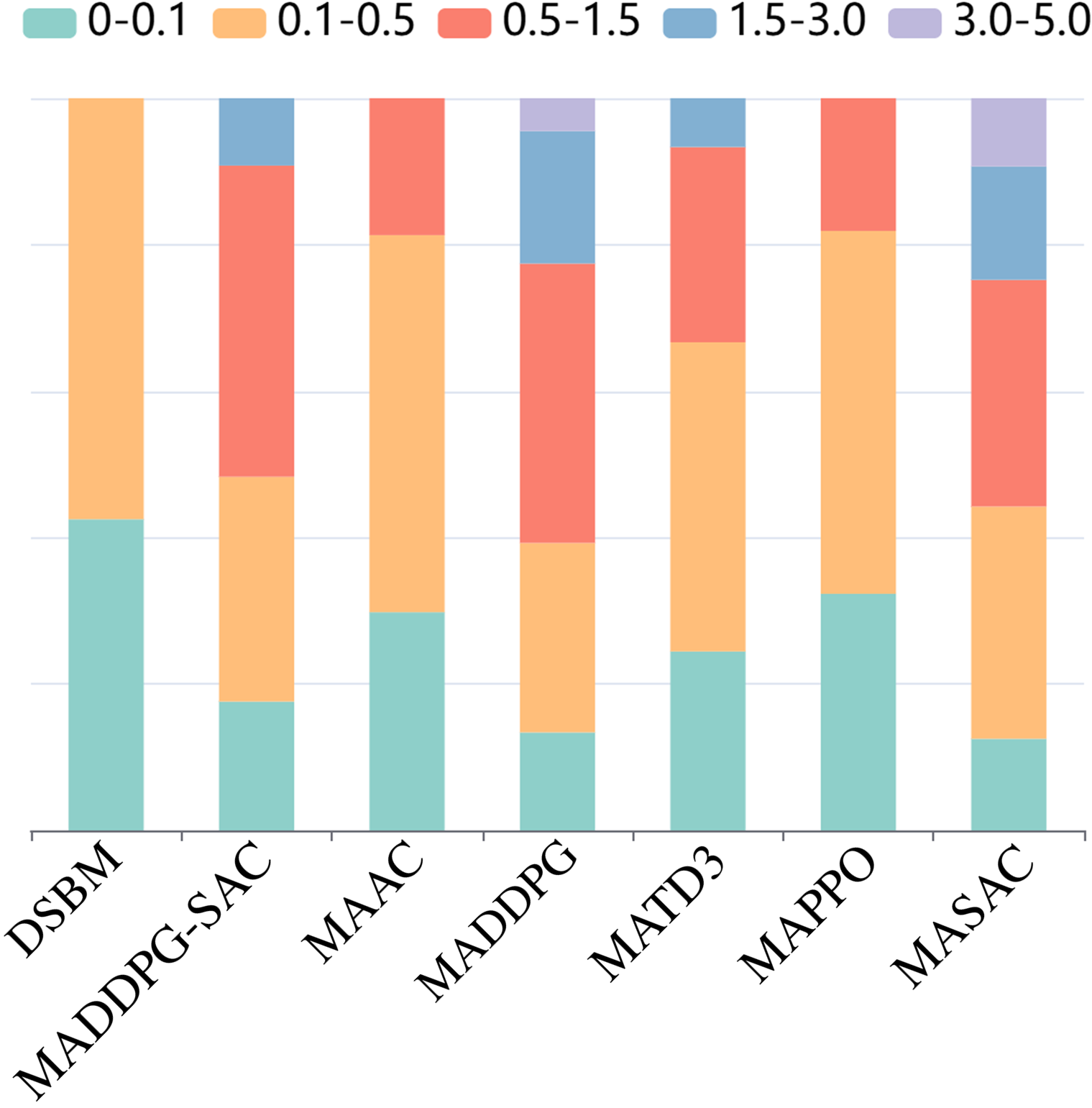}\label{fig4}}
	\hfil
	\subfloat[12 ET-AUVs tracking 4 targets with ocean current interference]
	{\includegraphics[width=0.24\textwidth]{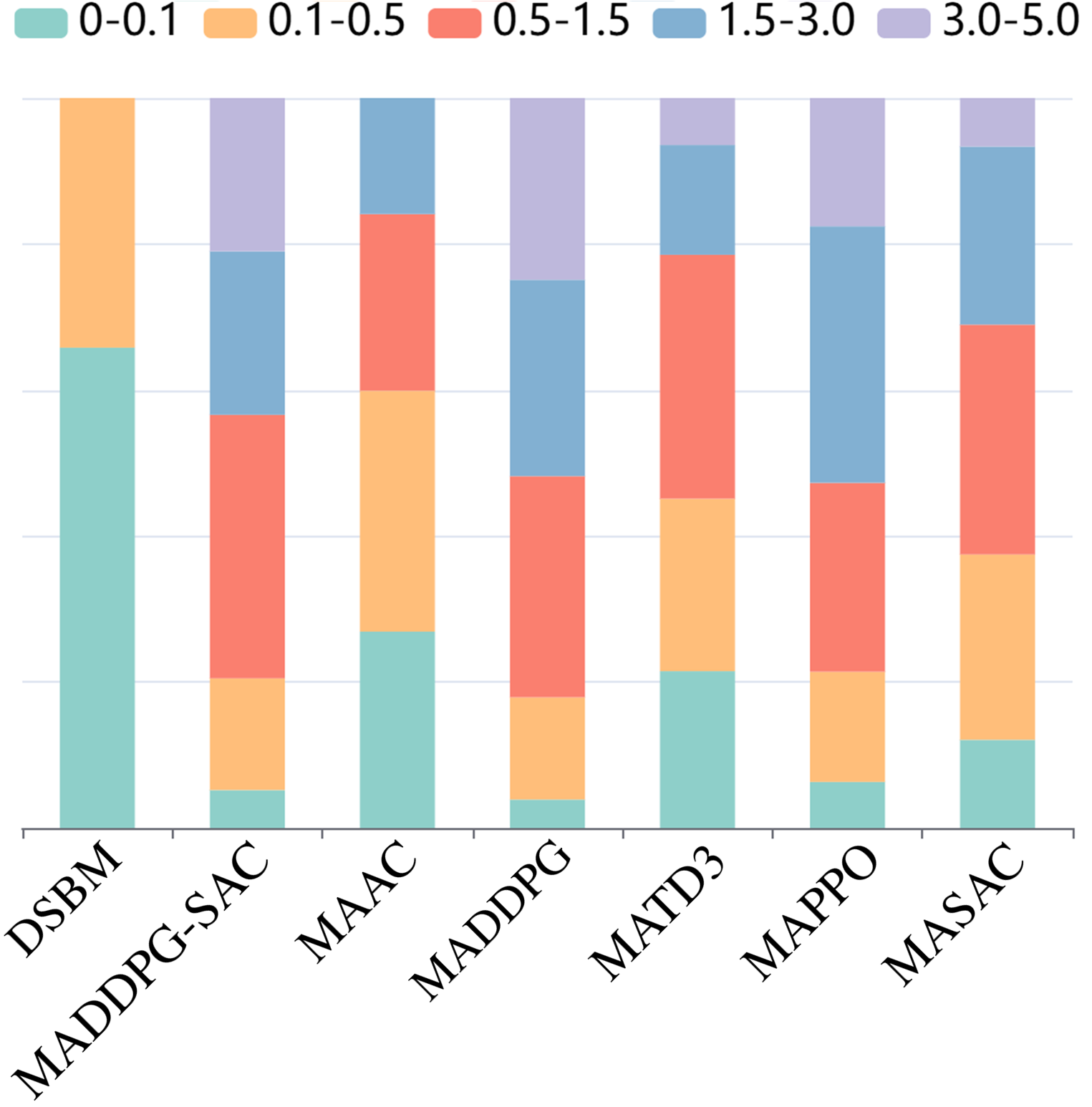}\label{fig5}}
 \caption{Distribution of AUV-Target distance comparison}
\end{figure}

\begin{figure}[bth]
	\centering 
	\subfloat[12 ET-AUVs tracking 4 targets without ocean current interference]
	{\includegraphics[width=0.24\textwidth]{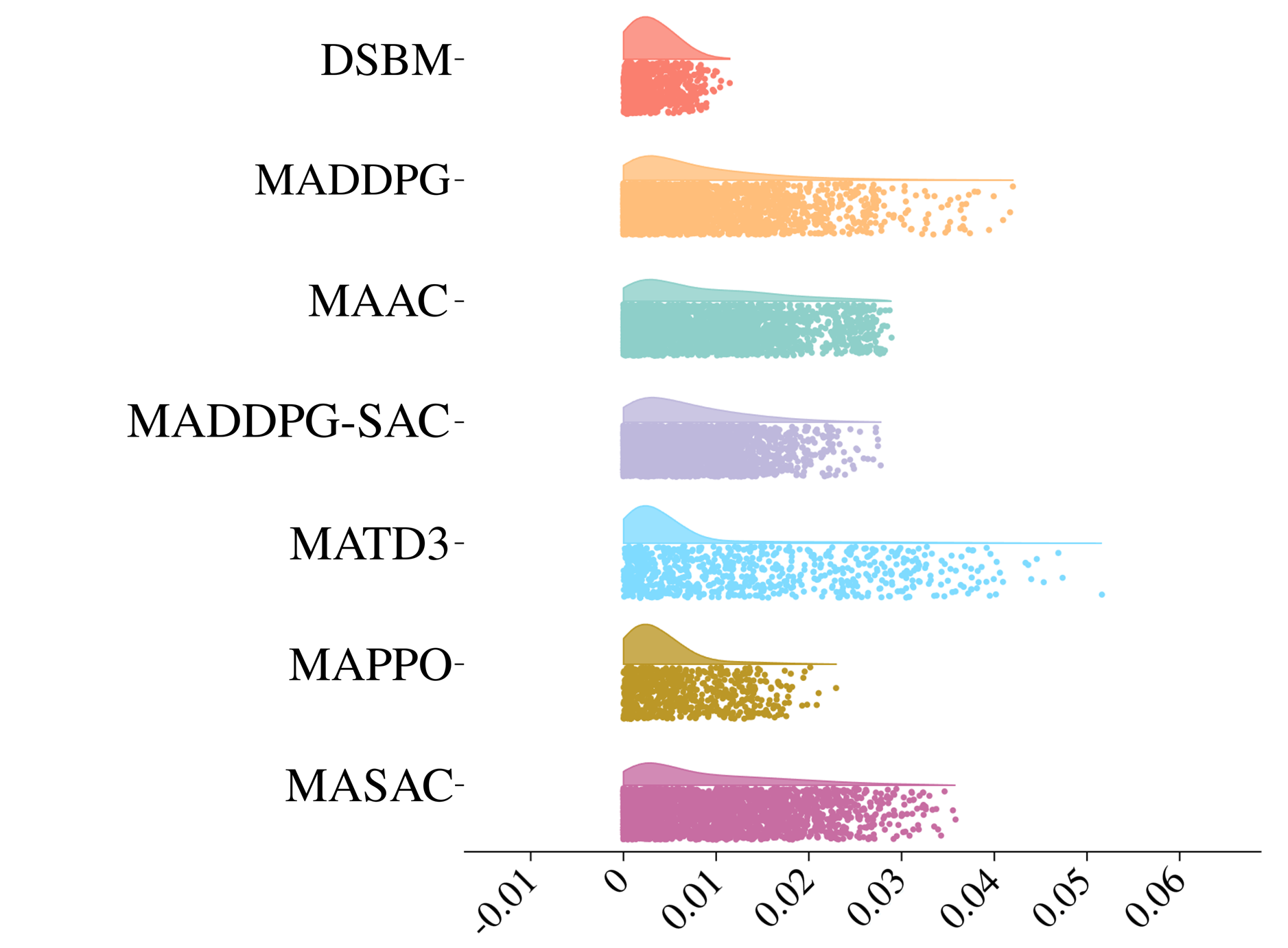}\label{fig6}}
	\hfil
	\subfloat[12 ET-AUVs tracking 4 targets with ocean current interference]
	{\includegraphics[width=0.24\textwidth]{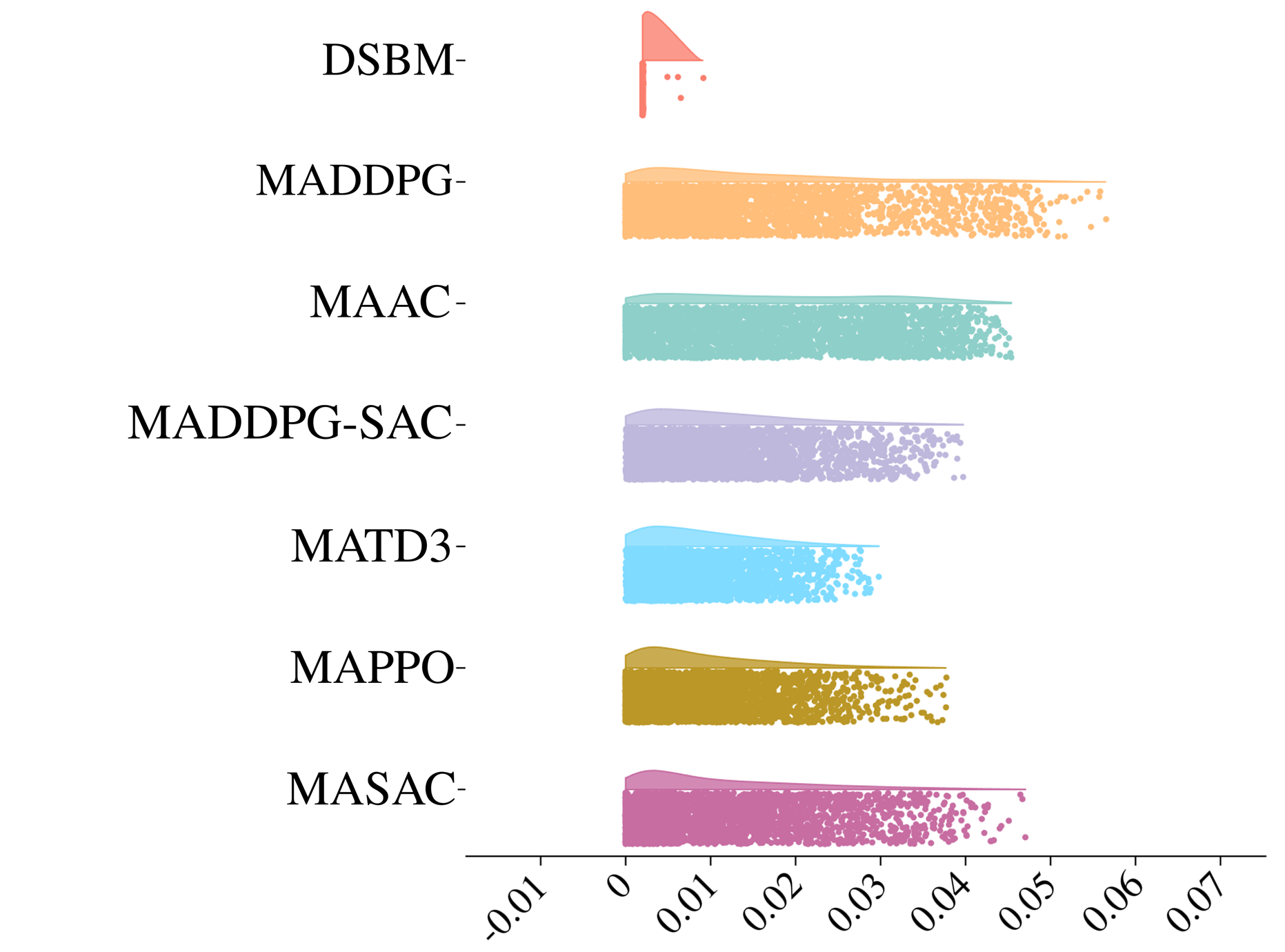}\label{fig7}}
 \caption{AUV-Target velocity difference comparison}
\end{figure}

\subsubsection{Tracking distance maintenance} 
As aforementioned, maintaining the distance between the ET-AUV and the target, i.e., the “AUV-target distance” is crucial during the tracking. Therefore, we also test the distribution of AUV-target distances with and without ocean current interference.

In conditions without ocean current interference, as shown in Fig. \ref{fig4}, the ET-AUV guided by DSBM almost always maintains a distance of less than 0.5 from the target, outperforming the other models. This is because our model quickly achieves a stable tracking state in the early stages which significantly shortens the initial capture process and reducing outliers in the AUV-target distance.

When ocean current interference is introduced, as shown in Fig. \ref{fig5}, the performance of the other six algorithms deteriorates to varying degrees, but the performance of DBSM improves. This is partly due to our reward reshaping term (Eq. \ref{EQ29}), which gives higher rewards to AUVs moving toward the target, guiding them to maintain the correct direction. 

Furthermore, the MAAC performs the second best, due to its layered attention mechanism. Specifically, the local attention layer in MAAC allows each AUV to focus on its own key information, while the global attention layer enables selective attention to important information from other AUVs. This means each AUV can effectively utilize other AUV’s information, ensuring each AUV maintains its distance from the target.}

\subsubsection{Velocity difference maintenance} 
In AUV swarm network-based underwater cooperative target tracking, frequent acceleration and deceleration for each ET-AUV can rapidly consume the energy, making against the long-duration tracking tasks especially in underwater environment. Therefore, the velocity difference between the ET-AUV and target, i.e., the “AUV-Target velocity difference” can be used to measure the endurance capacity and tracking efficiency of the tracking policy.

As shown in Fig. \ref{fig6}, the proposed DBSM performs the best with the smallest velocity difference without ocean current interference. MATD3 has the largest velocity difference, but over 80\% of the differences are within 0.005. This indicates that although the velocity of AUVs under MATD3 vary greatly, the overall velocity difference is also small. This is because MATD3 introduces noise to the policy network’s output, which helps the model explore more in the state-action space and avoid local optima. However, it also increases the randomness in action selection, causing the AUV to occasionally deviate from the intended state, leading to minor velocity differences.

As shown in Fig. \ref{fig7}, with ocean current interference, DBSM performs even better, with almost all data points close to the intended data (velocity). In contrast, the performance of the other algorithms declines to varying degrees. Although MATD3 shows a reduced maximum deviation, less than 40\% of the points fall within a 0.005 difference, indicating an overall decline in performance.

\begin{table*}[htbp]
\small
\centering
\caption{Strategic consistency comparison (12 ET-AUVs tracking 4 targets without ocean current interference)}
\label{consistency_without_current}
\begin{tabular}{lccccccc}
\hline
 & DSBM & MADDPG-SAC & MAAC & MADDPG & MATD3 & MAPPO & MASAC \\
\hline
All Different & 41.15\% & 46.23\% & 46.12\% & 51.87\% & 43.12\% & 40.27\% & 44.63\% \\
Two Alike & 25.34\% & 28.06\% & 33.52\% & 34.22\% & 30.97\% & 31.45\% & 36.61\% \\
All Alike & 33.51\% & 25.71\% & 20.36\% & 13.91\% & 25.91\% & 28.28\% & 18.76\% \\
\hline
\end{tabular}
\end{table*}

\begin{table*}[htbp]
\small
\centering
\caption{Strategic consistency comparison (12 ET-AUVs tracking 4 targets with ocean current interference)}
\label{consistency_with_current}
\begin{tabular}{lccccccc}
\hline
 & DSBM & MADDPG-SAC & MAAC & MADDPG & MATD3 & MAPPO & MASAC \\
\hline
All Different & 30.23\% & 48.79\% & 55.62\% & 53.10\% & 38.31\% & 50.12\% & 48.93\% \\
Two Alike & 18.25\% & 17.94\% & 17.38\% & 15.12\% & 24.11\% & 19.37\% & 26.98\% \\
All Alike & 51.52\% & 33.27\% & 27.00\% & 31.78\% & 37.58\% & 30.51\% & 24.09\% \\
\hline
\end{tabular}
\end{table*}

\subsubsection{Intra-formation strategy consistency} 
During the tracking process, a consistent strategy for the ET-AUV formation helps improve target identification and tracking efficiency, especially in complex underwater environments. This is crucial to prevent tracking errors or relief target loss due to environmental factors. 

We analyze the strategy consistency in a scenario where 12 ET-AUVs track 4 targets, with 3 ET-AUVs in each formation. There are three possible cases for strategy selections within each formation:

\begin{enumerate}
    \item \textbf{All Different:} Each ET-AUV in the formation has a different strategy.
    \item \textbf{Two Alike:} In the formation, there are two ET-AUVs whose strategies are the same.
    \item \textbf{All Alike:} The strategy is the same for all ET-AUVs in the formation.
\end{enumerate}

As shown in Table \ref{consistency_without_current}, the proportion of identical strategies within the DSBM-guided formation is the highest, indicating the best tracking performance of DSBM. This is because the formation under DSBM reaches a stable tracking state quickly. At this stage, the ET-AUVs and the target form a shape similar to a triangular pyramid, where the three ET-AUVs form the base triangle, and the target serves as the apex. In this situation, the target’s position appears nearly identical from each AUV’s perspective, leading to a convergence in strategy selection within the formation. Therefore, DSBM-guided formation proportions of exhibit the highest strategy consistency.

As shown in Table \ref{consistency_with_current}, when ocean current interference is applied, the proportion of strategy consistency increases for all the algorithms. This is because, in our design, the interference from ocean currents is similar within close position. This means that each ET-AUV within the same formation resists the current’s force, resulting in convergent action strategy selection. Additionally, our algorithm shows the greatest improvement and ultimately achieves the highest strategy consistency. The above phenomenon results from the following factors: 1) DSBM-guided formations can achieve a stable state more quickly; 2) our proposed Dynamic-Switching mechanism strengthens the ET-AUVs' resistance to ocean currents. Totally, both of the above factors result in consistent strategy selection when the current’s magnitude and direction are similar.

\subsubsection{Energy consumption} 
To visually demonstrate DSBM’s contribution to improving the energy efficiency for the tracking system, we test the system remaining energy during the training. The system energy consumption follows the following rules: 1) for target tracking, the ET-AUV has seven limited actions, i.e., up, down, left, right, forward, backward, and staying still; 2) energy is consumed only when the ET-AUV moves; 3) staying still consumes no energy. As shown in Fig. \ref{EnergyConsumptionComparisonWith} and Fig. \ref{EnergyConsumptionComparisonWithout}, DSBM achieves the most stable and efficient state. This is because our proposed DSBM encourages ET-AUVs to focus on velocity consistency. When the ocean current force aligns with the ET-AUV's movement direction, the ET-AUV is more likely to choose the staying still action.

\begin{figure}[htbp]
	\centering 
	\subfloat[12 ET-AUVs tracking 4 targets without ocean current interference]
	{\includegraphics[width=0.24\textwidth]{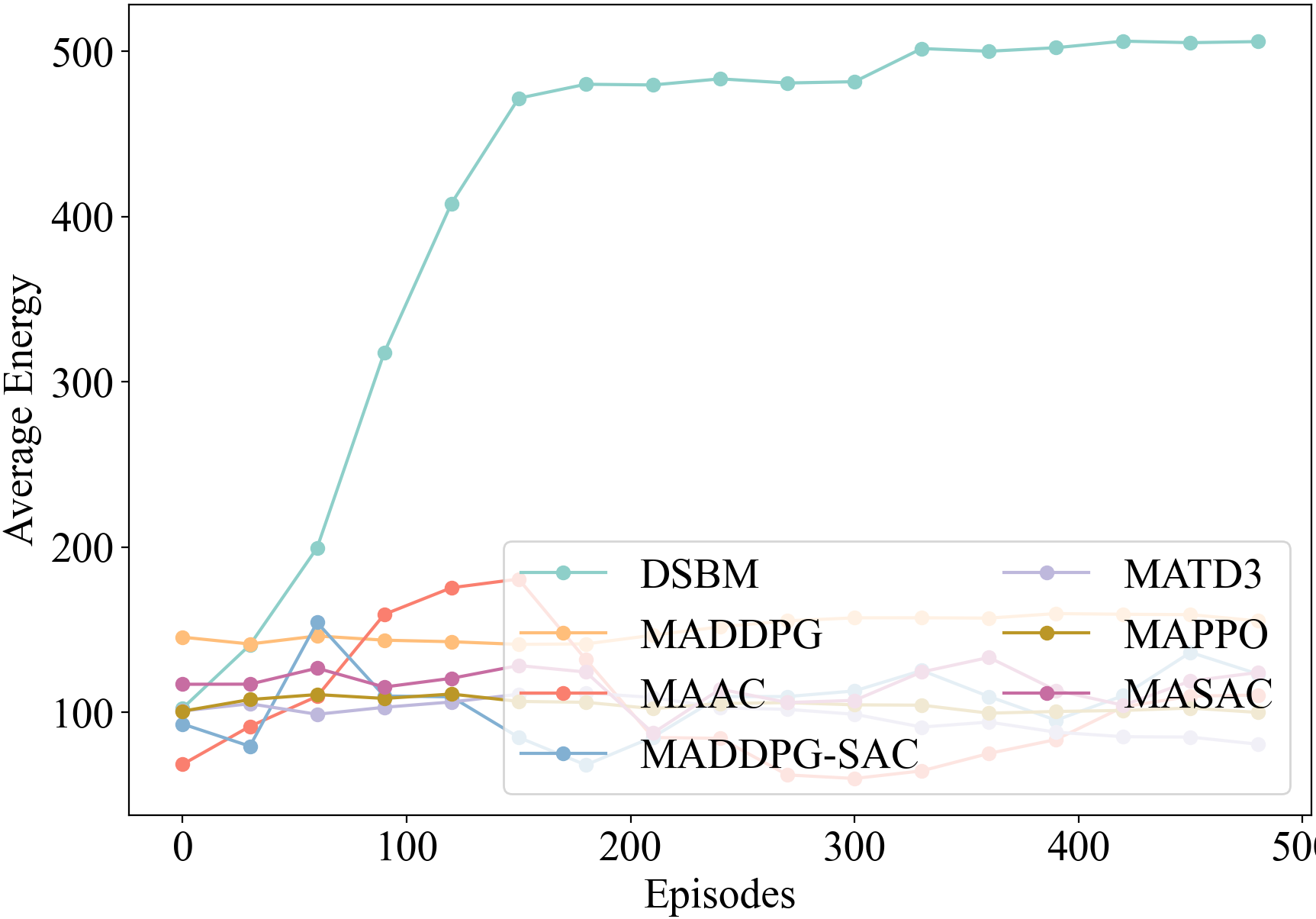}\label{EnergyConsumptionComparisonWith}}
	\hfil
	\subfloat[12 ET-AUVs tracking 4 targets with ocean current interference]
	{\includegraphics[width=0.24\textwidth]{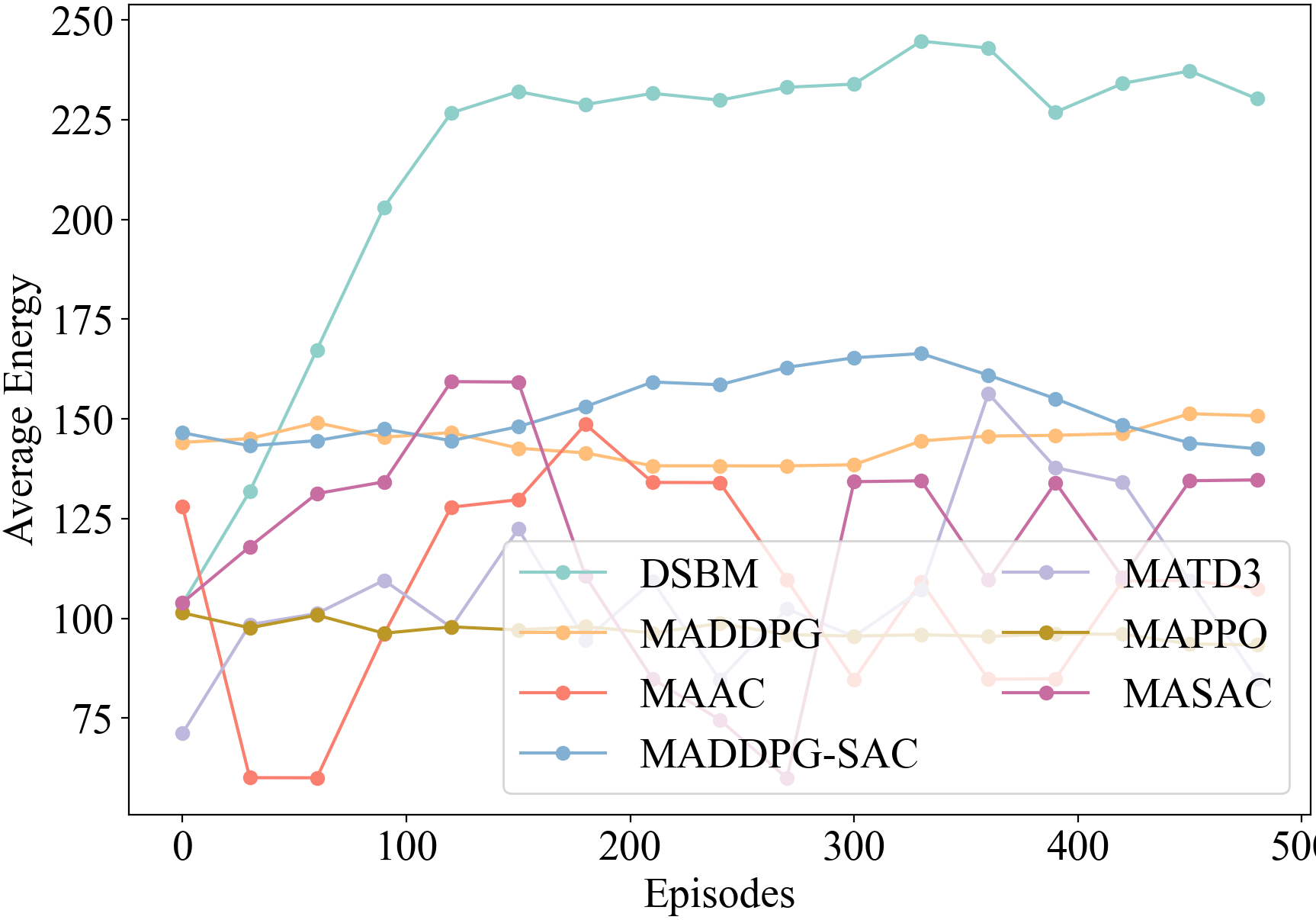}\label{EnergyConsumptionComparisonWithout}}
 \caption{Energy consumption comparison}
\end{figure}

\subsubsection{Tracking accuracy} 
In underwater cooperative tracking operations, tracking accuracy is a key factor in evaluating the model performance. We assume that if the actual distance between the ET-AUV and the target is within 0.005 of the preset distance, it counts as accurate tracking. 

As shown in Table \ref{tracking_accuracy}, with ocean current interference, DSBM achieves a tracking accuracy of 94.23\%. When the ocean current interference is introduced, DSBM’s tracking accuracy reaches 99.90\%. In both cases, DSBM performs the best.

\begin{table*}[htbp]
\small
\centering
\caption{Tracking accuracy comparison (12 ET-AUVs tracking 4 targets)}
\label{tracking_accuracy}
\begin{tabular}{lccccccc}
\hline
 & DSBM & MADDPG-SAC & MAAC & MADDPG & MATD3 & MAPPO & MASAC \\
\hline
Without ocean current interference & 94.23\% & 49.83\% & 55.40\% & 50.30\% & 83.30\% & 88.40\% & 46.83\% \\
With ocean current interference & 99.90\% & 28.77\% & 27.67\% & 26.77\% & 38.37\% & 41.97\% & 38.47\% \\
\hline
\end{tabular}
\end{table*}

\begin{figure*}[htbp]
  \centering
  \subfloat[Scenario 1\_Phase 1]{\includegraphics[width=0.3\textwidth]{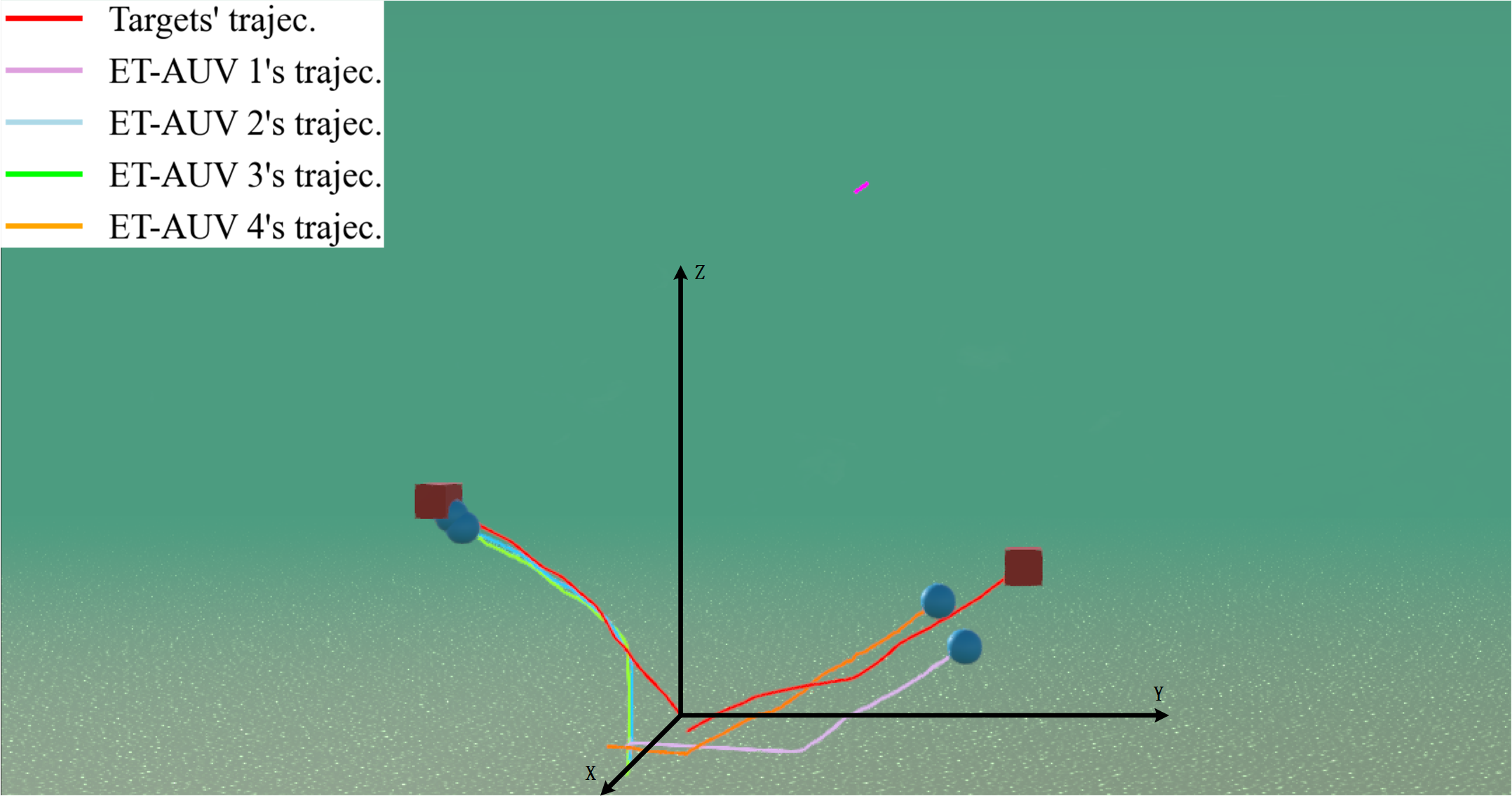}\label{f14-1-1}}\hfill
  \subfloat[Scenario 1\_Phase 2]{\includegraphics[width=0.3\textwidth]{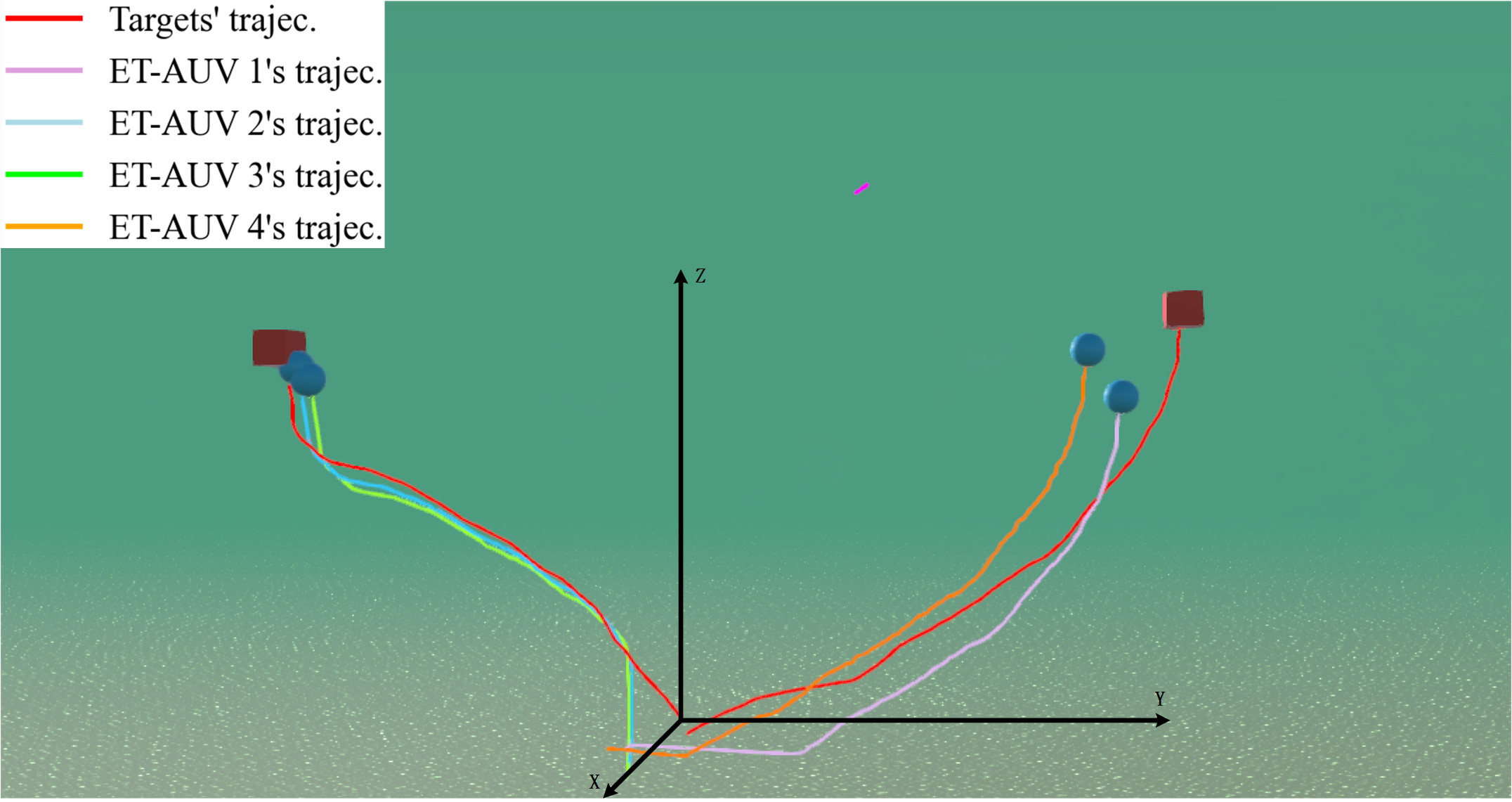}\label{f14-1-2}}\hfill
  \subfloat[Scenario 1\_Phase 3]{\includegraphics[width=0.3\textwidth]{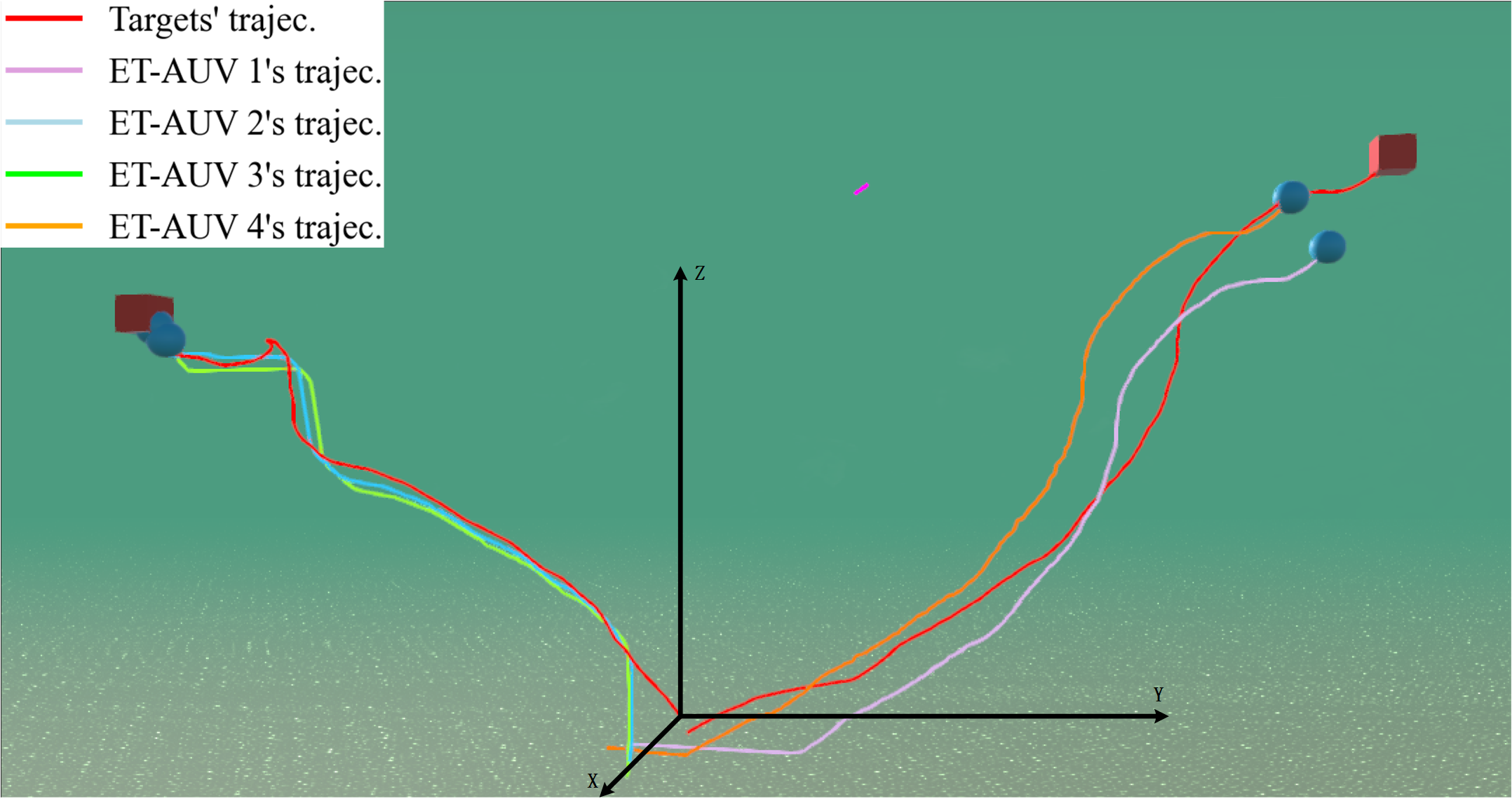}\label{f14-1-3}}
  
  \vspace{1ex}

  \subfloat[Scenario 2\_Phase 1]{\includegraphics[width=0.3\textwidth]{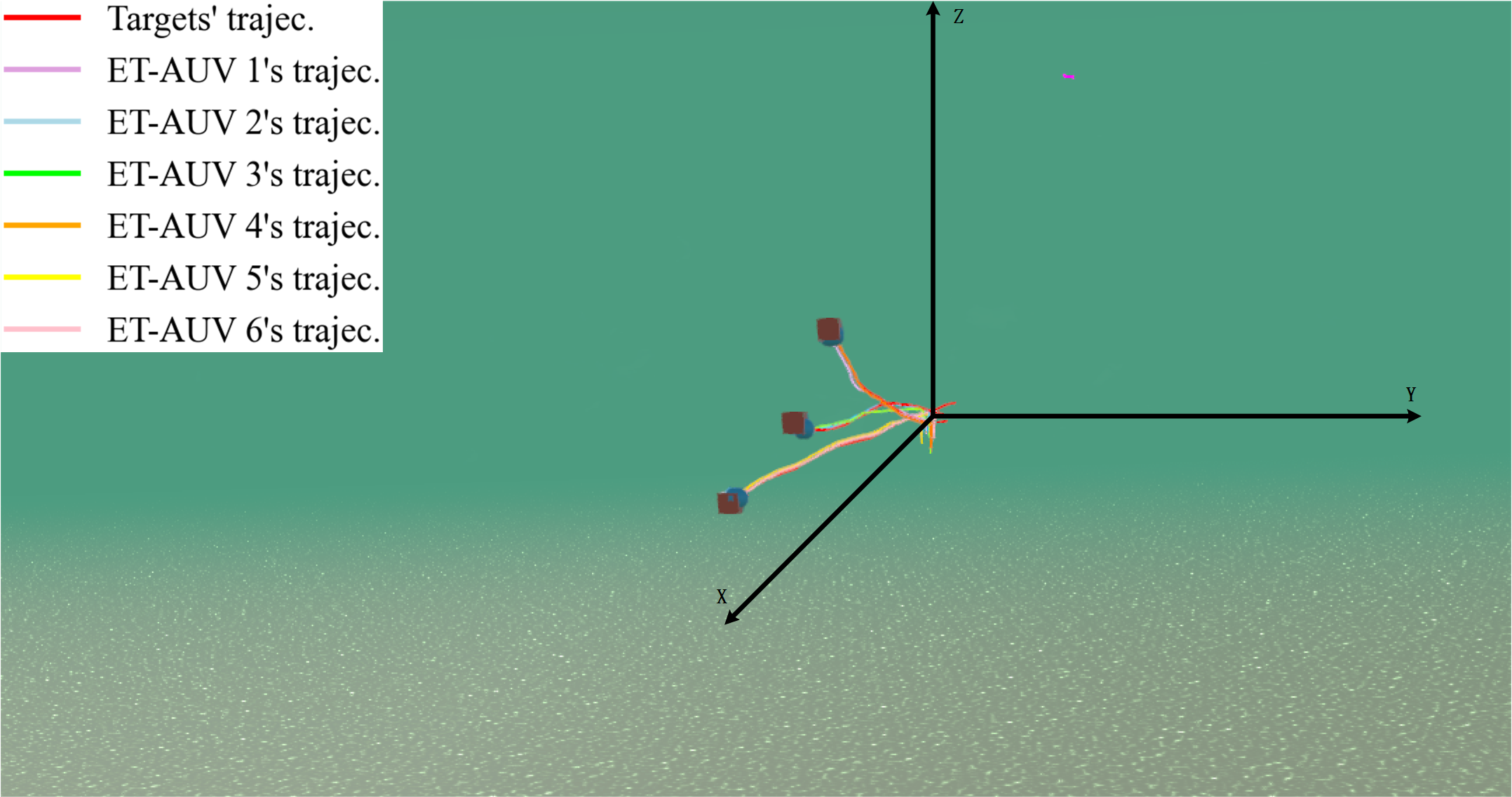}\label{f14-2-1}}\hfill
  \subfloat[Scenario 2\_Phase 2]{\includegraphics[width=0.3\textwidth]{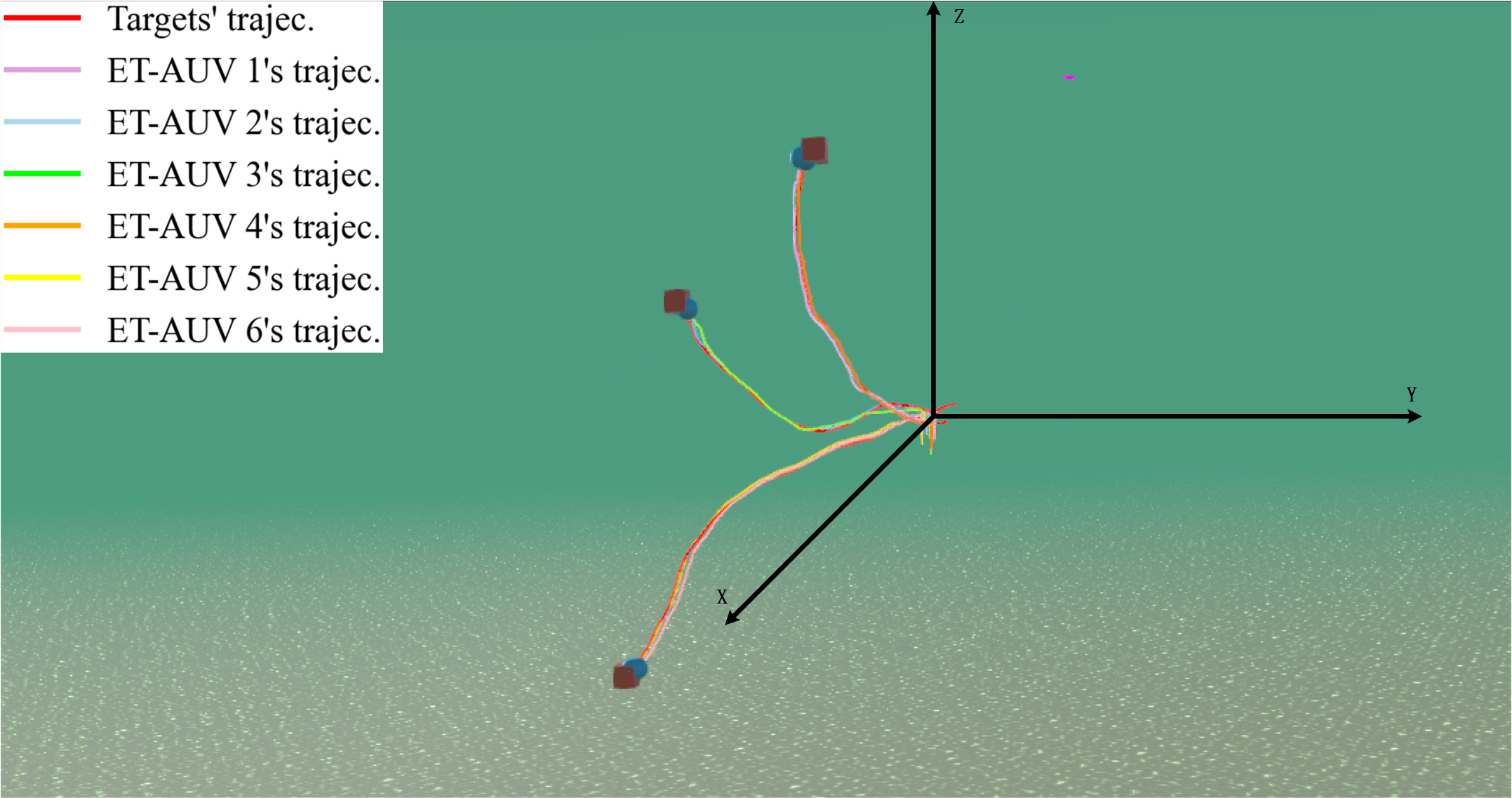}\label{f14-2-2}}\hfill
  \subfloat[Scenario 2\_Phase 3]{\includegraphics[width=0.3\textwidth]{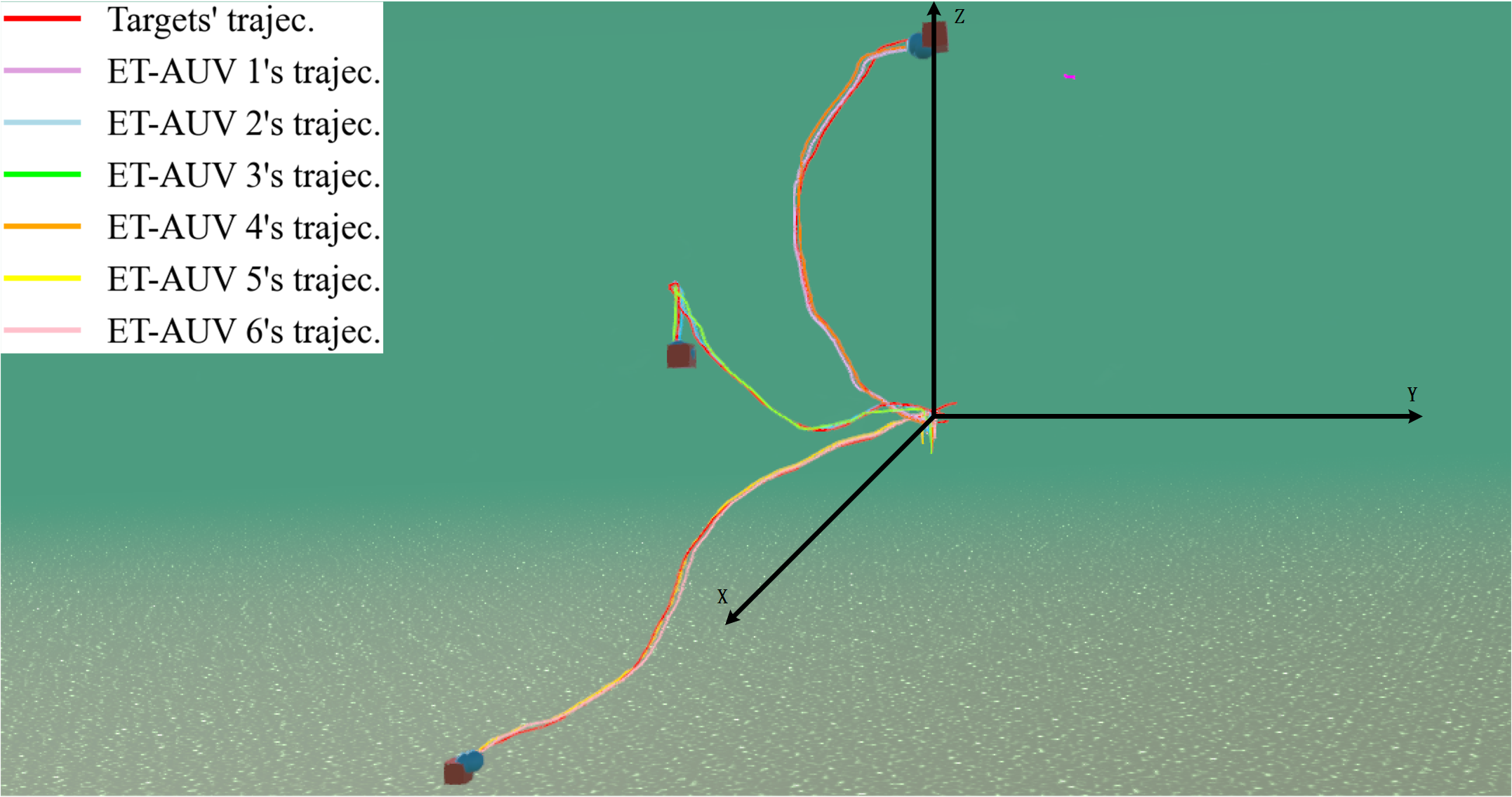}\label{f14-2-3}}

  \vspace{1ex}

  \subfloat[Scenario 3\_Phase 1]{\includegraphics[width=0.3\textwidth]{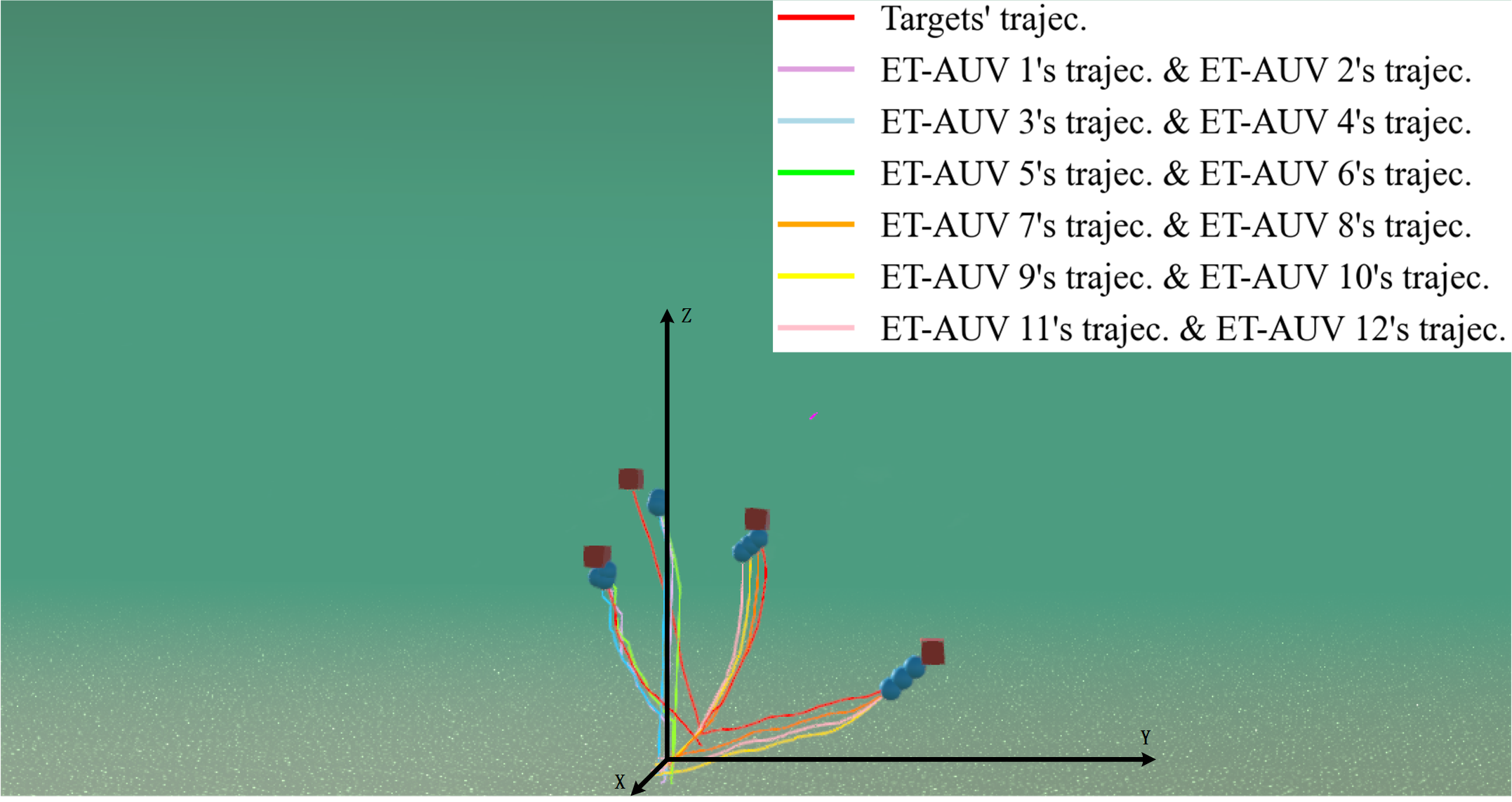}\label{f14-3-1}}\hfill
  \subfloat[Scenario 3\_Phase 2]{\includegraphics[width=0.3\textwidth]{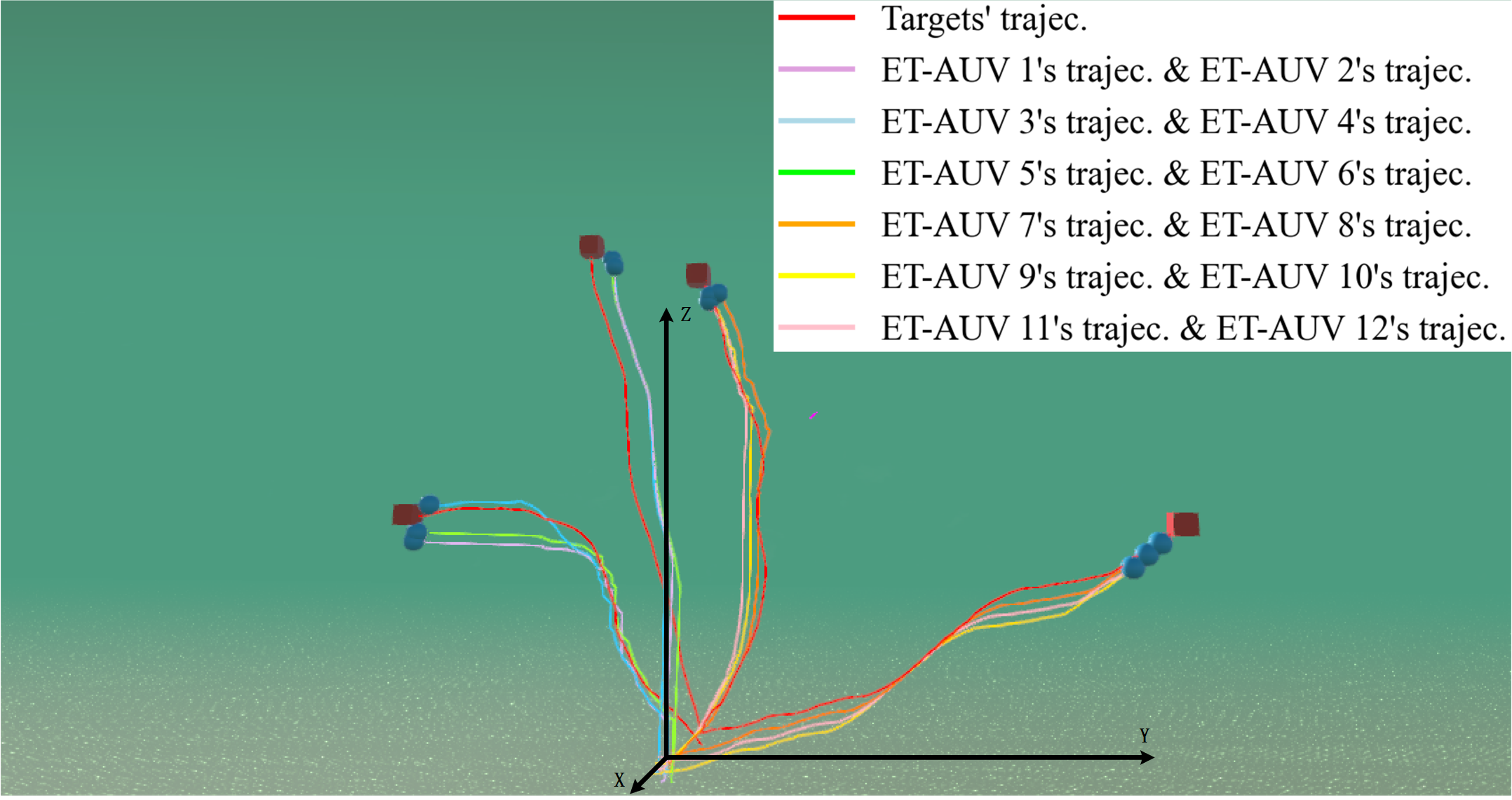}\label{f14-3-2}}\hfill
  \subfloat[Scenario 3\_Phase 3]{\includegraphics[width=0.3\textwidth]{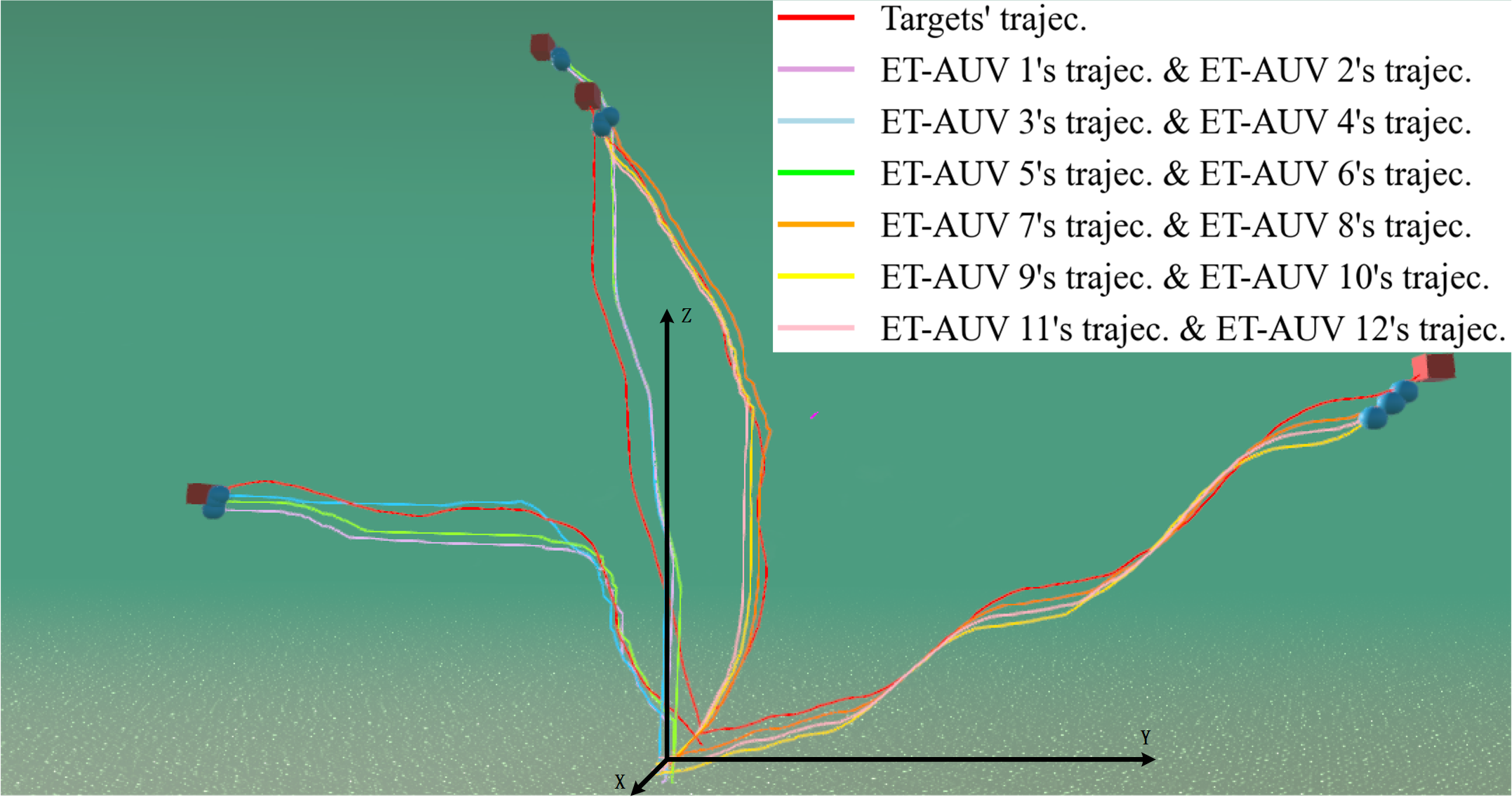}\label{f14-3-3}}
  
  \caption{Test for the availability of the proposed approaches in 3D environment (the word trajec. in the legend of each sub-figure is short for trajectory).}
  \label{fig8}
\end{figure*}

Finally, to make the entire tracking process directed by our proposed DSBM more clear, we use Unity \cite{Unity2024} to simulate the underwater environment for visualizing the tracking process.
We have embedded ocean currents and other marine elements into the MPE environment. 
It can realistically simulate our proposed method for multi-AUV tracking in underwater scenarios. 
Finally, we use Unity to visualize the trajectory data of each ET-AUV and the target.
As shown in Fig. \ref{fig8}, the red cubes represent the targets, and the blue spheres represent ET-AUVs. 
The red solid line represents the movement trajectory of the target, while other colored solid lines represent the tracking trajectories of the ET-AUVs. 
Specifically, we simulate three scenarios of multi-target tracking. 
Fig. \ref{f14-1-1}-Fig. \ref{f14-1-3} represent 4 ET-AUVs tracking 2 targets, Fig. \ref{f14-2-1}-Fig. \ref{f14-2-3} represent 6 ET-AUVs tracking 3 targets, and Fig. \ref{f14-3-1}-Fig. \ref{f14-3-3} represent 12 ET-AUVs tracking 4 targets. 
The visualization in Fig. \ref{fig8} also proves the effectiveness of the proposed DSBM in tracking underwater targets in a 3D environment.

\section{Conclusion}\label{Section:7}

In this paper, we take a deep study on multi-AUV cooperative underwater multi-target tracking.
We regard the AUV swarm as an underwater ad-hoc network and employ SDN technique to build HSARL architecture, a software-defined multi-AUV reinforcement learning architecture.
Based on HSARL, we propose the Dynamic-Switching-based MARL (DSBM)-driven tracking algorithm. It includes "Dynamic-Switching Attention" and "Dynamic-Switching Resampling" mechanisms to improve the AUV swarm network's self-learning efficiency and enhance tracking accuracy.
Additionally, to further expedite the convergence of the HSARL algorithm, especially during the early learning phase, we introduce a reward reshaping mechanism to ensure more rapid and stable learning outcomes.
Finally, we propose ASMA, an advanced classification algorithm to efficiently assigns formations in the presence of ocean current interference.
Evaluation results demonstrate that our proposed tracking algorithm achieves the fastest convergence speed and tracking performance compared to various popular research products. 
Future research directions derived from this work can be summarized as the following: 1) Optimizing underwater obstacle avoidance mechanisms for ET-AUVs to mitigate potential damage; 2) Balancing energy consumption among the AUVs to enhance the endurance of the AUV swarm network-based cooperative tracking systems; 3) Designing AUV swarm robustness control framework when the unstable underwater communication (e.g., the underwater acoustic-based communication) has to been taken into account.

\section{ACKNOWLEDGEMENTS}

The work is supported by the Joint Funds of the National Natural Science Foundation of China (No. U22A2011), Frontier Technologies R\&D Program of Jiangsu (No. BF2024072), and National Training Program of Innovation and Entrepreneurship for Undergraduates (No. 202410145004). 
\ifCLASSOPTIONcaptionsoff
  \newpage
\fi



%

\bibliographystyle{IEEEtran}
\bibliography{ref}

\begin{thebibliography}{10}
\providecommand{\url}[1]{#1}
\csname url@samestyle\endcsname
\providecommand{\newblock}{\relax}
\providecommand{\bibinfo}[2]{#2}
\providecommand{\BIBentrySTDinterwordspacing}{\spaceskip=0pt\relax}
\providecommand{\BIBentryALTinterwordstretchfactor}{4}
\providecommand{\BIBentryALTinterwordspacing}{\spaceskip=\fontdimen2\font plus
\BIBentryALTinterwordstretchfactor\fontdimen3\font minus \fontdimen4\font\relax}
\providecommand{\BIBforeignlanguage}[2]{{%
\expandafter\ifx\csname l@#1\endcsname\relax
\typeout{** WARNING: IEEEtran.bst: No hyphenation pattern has been}%
\typeout{** loaded for the language `#1'. Using the pattern for}%
\typeout{** the default language instead.}%
\else
\language=\csname l@#1\endcsname
\fi
#2}}
\providecommand{\BIBdecl}{\relax}
\BIBdecl

\bibitem{9633006}
F.~A. Setiawan and P.~Rahmadi, ``Indoalgae: The database of indonesian native strains of potential marine algae,'' in \emph{2021 Sixth International Conference on Informatics and Computing (ICIC)}, 2021, pp. 1--5.

\bibitem{10244971}
Z.~Yan, ``Revolution system of spatial-temporal pattern of marine resources development and management based on deep learning algorithms,'' in \emph{2023 International Conference on Data Science and Network Security (ICDSNS)}, 2023, pp. 1--6.

\bibitem{9707741}
L.~Zong, H.~Wang, and G.~Luo, ``Transmission control over satellite network for marine environmental monitoring system,'' \emph{IEEE Transactions on Intelligent Transportation Systems}, vol.~23, no.~10, pp. 19\,668--19\,675, 2022.

\bibitem{159741}
S.~Yuan, Y.~Li, F.~Bao, H.~Xu, Y.~Yang, Q.~Yan, S.~Zhong, H.~Yin, J.~Xu, Z.~Huang, and J.~Lin, ``Marine environmental monitoring with unmanned vehicle platforms: Present applications and future prospects,'' \emph{Science of The Total Environment}, vol. 858, no. Part 1, p. 159741, 2023.

\bibitem{9618770}
S.~Hou, W.~Li, T.~Liu, S.~Zhou, J.~Guan, R.~Qin, and Z.~Wang, ``D2cl: A dense dilated convolutional lstm model for sea surface temperature prediction,'' \emph{IEEE Journal of Selected Topics in Applied Earth Observations and Remote Sensing}, vol.~14, pp. 12\,514--12\,523, 2021.

\bibitem{112178}
Q.~Ma, D.~Zhang, C.~Wan, J.~Zhang, and N.~Lyu, ``Multi-objective emergency resources allocation optimization for maritime search and rescue considering accident black-spots,'' \emph{Ocean Engineering}, vol. 261, p. 112178, 2022.

\bibitem{Wang2023}
X.~Wang, B.~Xu, and Y.~Guo, ``Fuzzy logic system-based robust adaptive control of auv with target tracking,'' \emph{International Journal of Fuzzy Systems}, vol.~25, no.~1, pp. 338--346, 2023.

\bibitem{9485055}
G.~Han, A.~Gong, H.~Wang, M.~Martínez-García, and Y.~Peng, ``Multi-auv collaborative data collection algorithm based on q-learning in underwater acoustic sensor networks,'' \emph{IEEE Transactions on Vehicular Technology}, vol.~70, no.~9, pp. 9294--9305, 2021.

\bibitem{9508160}
Z.~Yan, K.~Zhang, L.~Qiao, Y.~Hu, and B.~Song, ``A multiload wireless power transfer system with concentrated magnetic field for auv cluster system,'' \emph{IEEE Transactions on Industry Applications}, vol.~58, no.~1, pp. 1307--1314, 2022.

\bibitem{10329958}
B.~Jiang, J.~Du, C.~Jiang, Z.~Han, and M.~Debbah, ``Underwater searching and multiround data collection via auv swarms: An energy-efficient aoi-aware mappo approach,'' \emph{IEEE Internet of Things Journal}, vol.~11, no.~7, pp. 12\,768--12\,782, 2024.

\bibitem{9530372}
N.~T. Hung, F.~F.~C. Rego, and A.~M. Pascoal, ``Cooperative distributed estimation and control of multiple autonomous vehicles for range-based underwater target localization and pursuit,'' \emph{IEEE Transactions on Control Systems Technology}, vol.~30, no.~4, pp. 1433--1447, 2022.

\bibitem{10214407}
T.~Zhou, Y.~Wang, L.~Zhang, B.~Chen, and X.~Yu, ``Underwater multitarget tracking method based on threshold segmentation,'' \emph{IEEE Journal of Oceanic Engineering}, vol.~48, no.~4, pp. 1255--1269, 2023.

\bibitem{9390453}
H.~Flores, N.~H. Motlagh, A.~Zuniga, M.~Liyanage, M.~Passananti, S.~Tarkoma, M.~Youssef, and P.~Nurmi, ``Toward large-scale autonomous marine pollution monitoring,'' \emph{IEEE Internet of Things Magazine}, vol.~4, no.~1, pp. 40--45, 2021.

\bibitem{10161282}
Y.~Girdhar, N.~McGuire, L.~Cai, S.~Jamieson, S.~McCammon, B.~Claus, J.~E.~S. Soucie, J.~E. Todd, and T.~A. Mooney, ``Curee: A curious underwater robot for ecosystem exploration,'' in \emph{2023 IEEE International Conference on Robotics and Automation (ICRA)}, 2023, pp. 11\,411--11\,417.

\bibitem{9681175}
S.~Rani, H.~Babbar, P.~Kaur, M.~D. Alshehri, and S.~H. Shah, ``An optimized approach of dynamic target nodes in wireless sensor network using bio inspired algorithms for maritime rescue,'' \emph{IEEE Transactions on Intelligent Transportation Systems}, vol.~24, no.~2, pp. 2548--2555, 2023.

\bibitem{9532493}
G.~Soldi, D.~Gaglione, N.~Forti, A.~D. Simone, F.~C. Daffinà, G.~Bottini, D.~Quattrociocchi, L.~M. Millefiori, P.~Braca, S.~Carniel, P.~Willett, A.~Iodice, D.~Riccio, and A.~Farina, ``Space-based global maritime surveillance. part i: Satellite technologies,'' \emph{IEEE Aerospace and Electronic Systems Magazine}, vol.~36, no.~9, pp. 8--28, 2021.

\bibitem{9722969}
Z.~Chu, F.~Wang, T.~Lei, and C.~Luo, ``Path planning based on deep reinforcement learning for autonomous underwater vehicles under ocean current disturbance,'' \emph{IEEE Transactions on Intelligent Vehicles}, vol.~8, no.~1, pp. 108--120, 2023.

\bibitem{9625977}
W.~Lan, X.~Jin, T.~Wang, and H.~Zhou, ``Improved rrt algorithms to solve path planning of multi-glider in time-varying ocean currents,'' \emph{IEEE Access}, vol.~9, pp. 158\,098--158\,115, 2021.

\bibitem{9356608}
Y.~Yang, Y.~Xiao, and T.~Li, ``A survey of autonomous underwater vehicle formation: Performance, formation control, and communication capability,'' \emph{IEEE Communications Surveys \& Tutorials}, vol.~23, no.~2, pp. 815--841, 2021.

\bibitem{9451536}
Z.~Fang, J.~Wang, J.~Du, X.~Hou, Y.~Ren, and Z.~Han, ``Stochastic optimization-aided energy-efficient information collection in internet of underwater things networks,'' \emph{IEEE Internet of Things Journal}, vol.~9, no.~3, pp. 1775--1789, 2022.

\bibitem{9332276}
H.~Guo, Z.~Sun, and P.~Wang, ``Joint design of communication, wireless energy transfer, and control for swarm autonomous underwater vehicles,'' \emph{IEEE Transactions on Vehicular Technology}, vol.~70, no.~2, pp. 1821--1835, 2021.

\bibitem{9498989}
J.~Tang, G.~Liu, and Q.~Pan, ``A review on representative swarm intelligence algorithms for solving optimization problems: Applications and trends,'' \emph{IEEE/CAA Journal of Automatica Sinica}, vol.~8, no.~10, pp. 1627--1643, 2021.

\bibitem{10004961}
C.~Lin, G.~Han, J.~Jiang, C.~Li, S.~B.~H. Shah, and Q.~Liu, ``Underwater pollution tracking based on software-defined multi-tier edge computing in 6g-based underwater wireless networks,'' \emph{IEEE Journal on Selected Areas in Communications}, vol.~41, no.~2, pp. 491--503, 2023.

\bibitem{9343333}
C.~Lin, G.~Han, T.~Wang, Y.~Bi, J.~Du, and B.~Zhang, ``Fast node clustering based on an improved birch algorithm for data collection towards software-defined underwater acoustic sensor networks,'' \emph{IEEE Sensors Journal}, vol.~21, no.~22, pp. 25\,480--25\,488, 2021.

\bibitem{9750063}
C.~Lin, G.~Han, T.~Zhang, S.~B.~H. Shah, and Y.~Peng, ``Smart underwater pollution detection based on graph-based multi-agent reinforcement learning towards auv-based network its,'' \emph{IEEE Transactions on Intelligent Transportation Systems}, vol.~24, no.~7, pp. 7494--7505, 2023.

\bibitem{9205627}
F.~Li, H.~Yao, J.~Du, C.~Jiang, Z.~Han, and Y.~Liu, ``Auction design for edge computation offloading in sdn-based ultra dense networks,'' \emph{IEEE Transactions on Mobile Computing}, vol.~21, no.~5, pp. 1580--1595, 2022.

\bibitem{9634122}
D.~M. Casas-Velasco, O.~M.~C. Rendon, and N.~L.~S. da~Fonseca, ``Drsir: A deep reinforcement learning approach for routing in software-defined networking,'' \emph{IEEE Transactions on Network and Service Management}, vol.~19, no.~4, pp. 4807--4820, 2022.

\bibitem{9076113}
C.~Lin, G.~Han, M.~Guizani, Y.~Bi, J.~Du, and L.~Shu, ``An sdn architecture for auv-based underwater wireless networks to enable cooperative underwater search,'' \emph{IEEE Wireless Communications}, vol.~27, no.~3, pp. 132--139, 2020.

\bibitem{HUY2023113202}
D.~Q. Huy, N.~Sadjoli, A.~B. Azam, B.~Elhadidi, Y.~Cai, and G.~Seet, ``Object perception in underwater environments: a survey on sensors and sensing methodologies,'' \emph{Ocean Engineering}, vol. 267, p. 113202, 2023.

\bibitem{CHEN2021109073}
G.~Chen, Y.~Shen, N.~Qu, and B.~He, ``Path planning of auv during diving process based on behavioral decision-making,'' \emph{Ocean Engineering}, vol. 234, p. 109073, 2021.

\bibitem{FANG2022110452}
Y.~Fang, Z.~Huang, J.~Pu, and J.~Zhang, ``Auv position tracking and trajectory control based on fast-deployed deep reinforcement learning method,'' \emph{Ocean Engineering}, vol. 245, p. 110452, 2022.

\bibitem{9685323}
Z.~Yang, J.~Du, Z.~Xia, C.~Jiang, A.~Benslimane, and Y.~Ren, ``Secure and cooperative target tracking via auv swarm: A reinforcement learning approach,'' in \emph{2021 IEEE Global Communications Conference (GLOBECOM)}, 2021, pp. 1--6.

\bibitem{110495}
Z.~Yan, C.~Zhang, W.~Tian, and M.~Zhang, ``Formation trajectory tracking control of discrete-time multi-auv in a weak communication environment,'' \emph{Ocean Engineering}, vol. 245, p. 110495, 2022.

\bibitem{9206115}
J.~Du, C.~Jiang, J.~Wang, Y.~Ren, and M.~Debbah, ``Machine learning for 6g wireless networks: Carrying forward enhanced bandwidth, massive access, and ultrareliable/low-latency service,'' \emph{IEEE Vehicular Technology Magazine}, vol.~15, no.~4, pp. 122--134, 2020.

\bibitem{10008249}
K.~G. Omeke, M.~Mollel, S.~T. Shah, K.~Arshad, L.~Zhang, Q.~H. Abbasi, and M.~A. Imran, ``Dynamic clustering and data aggregation for the internet-of-underwater-things networks,'' in \emph{2022 14th International Conference on Computational Intelligence and Communication Networks (CICN)}, 2022, pp. 322--328.

\bibitem{mordatch2017emergence}
I.~Mordatch and P.~Abbeel, ``Emergence of grounded compositional language in multi-agent populations,'' \emph{arXiv preprint arXiv:1703.04908}, 2017.

\bibitem{37}
R.~Lowe, Y.~Wu, A.~Tamar, J.~Harb, P.~Abbeel, and I.~Mordatch, ``Multi-agent actor-critic for mixed cooperative-competitive environments,'' \emph{Neural Information Processing Systems (NIPS)}, 2017.

\bibitem{38}
S.~Iqbal and F.~Sha, ``Actor-attention-critic for multi-agent reinforcement learning,'' in \emph{International conference on machine learning}, vol.~97.\hskip 1em plus 0.5em minus 0.4em\relax PMLR, 2019, pp. 2961--2970.

\bibitem{39}
H.~Yong, J.~Seo, J.~Kim, M.~Kim, and J.~Choi, ``Suspension control strategies using switched soft actor-critic models for real roads,'' \emph{IEEE Transactions on Industrial Electronics}, vol.~70, no.~1, pp. 824--832, Jan 2023.

\bibitem{ackermann2019reducing}
J.~Ackermann, V.~Gabler, T.~Osa, and M.~Sugiyama, ``Reducing overestimation bias in multi-agent domains using double centralized critics,'' 2019.

\bibitem{yu2022the}
C.~Yu, A.~Velu, E.~Vinitsky, J.~Gao, Y.~Wang, A.~Bayen, and Y.~Wu, ``The surprising effectiveness of {PPO} in cooperative multi-agent games,'' in \emph{Thirty-sixth Conference on Neural Information Processing Systems Datasets and Benchmarks Track}, 2022.

\bibitem{9446746}
X.~Wu, X.~Li, J.~Li, P.~C. Ching, V.~C.~M. Leung, and H.~V. Poor, ``Caching transient content for iot sensing: Multi-agent soft actor-critic,'' \emph{IEEE Transactions on Communications}, vol.~69, no.~9, pp. 5886--5901, 2021.

\bibitem{Unity2024}
U.~Technologies, ``Unity user manual (2024.1 lts),'' \url{https://docs.unity3d.com/Manual/index.html}, 2024, september 4, 2024.

\end{thebibliography}
\vspace{2ex}
\begin{IEEEbiography}[{\includegraphics[width=1in,height=1.25in,clip,keepaspectratio]{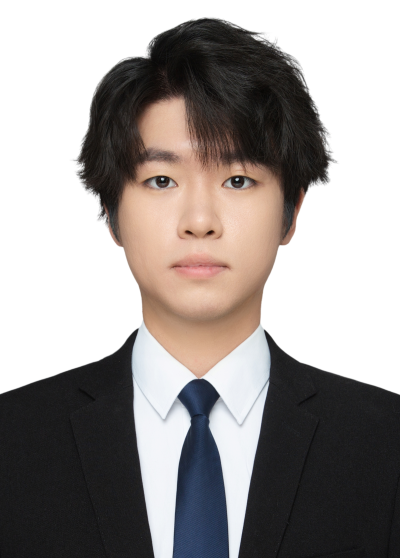}}]{Shengbo Wang}
is currently pursuing a Bachelor's degree at the Software College, Northeastern University, Shenyang, China. His current research interests include reinforcement learning, computer vision, and machine learning.
\end{IEEEbiography}

\vspace{2ex}
\begin{IEEEbiography}[{\includegraphics[width=1in,height=1.25in,clip,keepaspectratio]{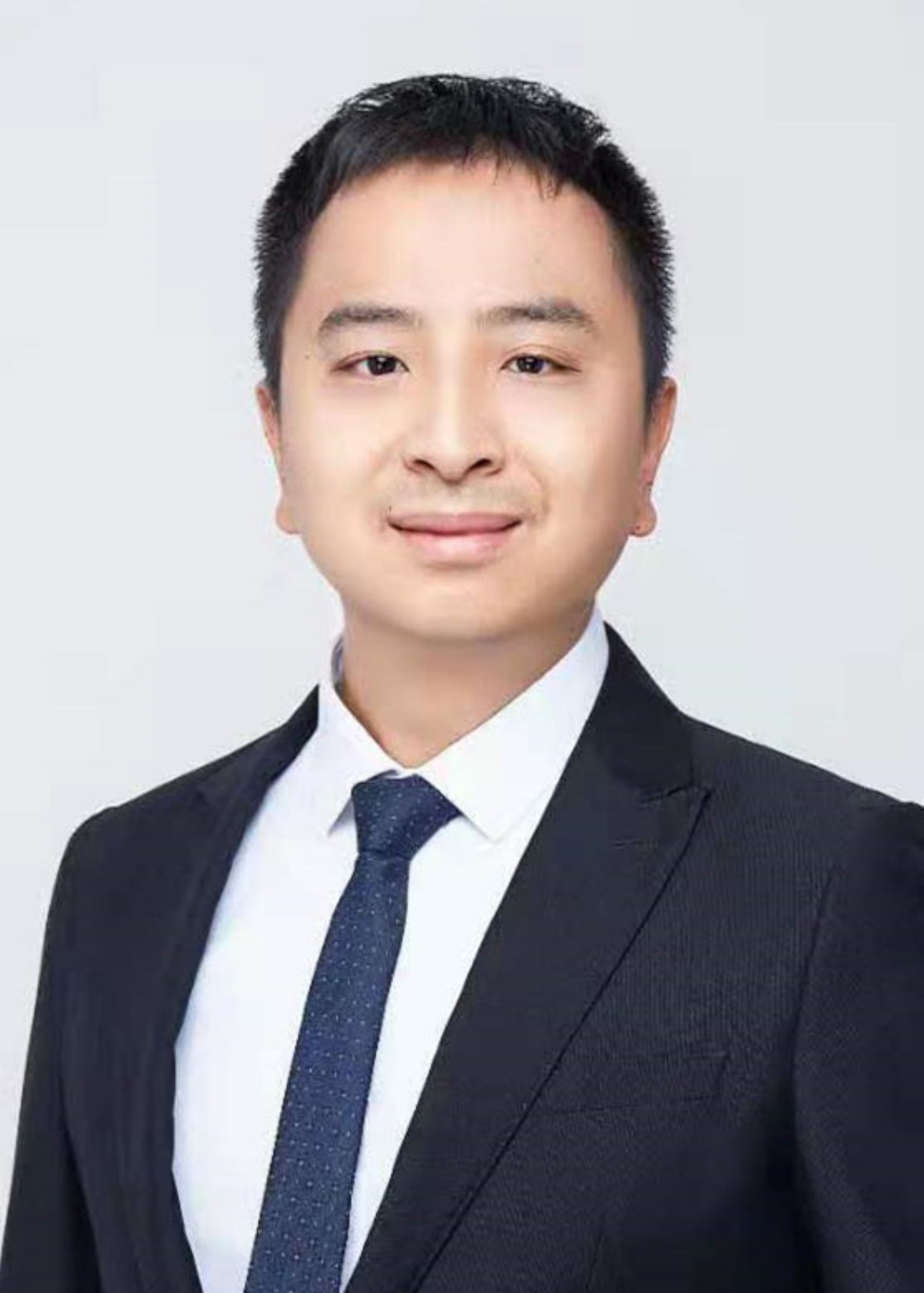}}]{Chuan Lin}
	[S'17, M'20] is currently an associate professor with the Software College, Northeastern University, Shenyang, China.
	He received the B.S. degree in Computer Science and Technology from Liaoning University, Shenyang, China in 2011, the M.S. degree in Computer Science and Technology from Northeastern University, Shenyang, China in 2013, and the Ph.D. degree in computer architecture in 2018.
	From Nove. 2018 to  Nove. 2020, he is a Postdoctoral Researcher with the School of Software, Dalian University of Technology, Dalian, China.
	His research interests include UWSNs, industrial IoT, software-defined networking.
\end{IEEEbiography}
\vspace{2ex}
\begin{IEEEbiography}[{\includegraphics[width=1in,height=1.25in,clip,keepaspectratio]{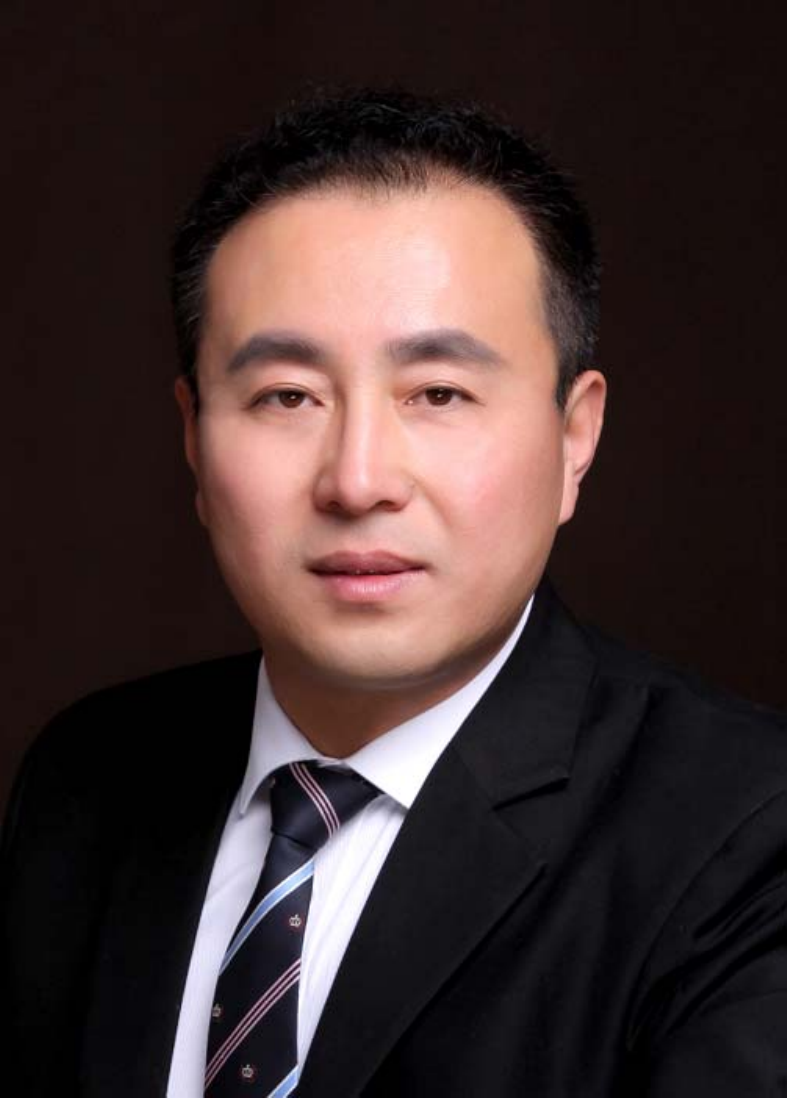}}]{Guangjie Han} [S’03-M’05-SM’18-F’22] 
is currently a Professor with the Department of Internet of Things Engineering, Hohai University, Changzhou, China. He received his Ph.D. degree from Northeastern University, Shenyang, China, in 2004. In February 2008, he finished his work as a Postdoctoral Researcher with the Department of Computer Science, Chonnam National University, Gwangju, Korea. From October 2010 to October 2011, he was a Visiting Research Scholar with Osaka University, Suita, Japan. From January 2017 to February 2017, he was a Visiting Professor with City University of Hong Kong, China. From July 2017 to July 2020, he was a Distinguished Professor with Dalian University of Technology, China. His current research interests include Internet of Things, Industrial Internet, Machine Learning and Artificial Intelligence, Mobile Computing, Security and Privacy. Dr. Han has over 500 peer-reviewed journal and conference papers, in addition to 160 granted and pending patents. Currently, his H-index is 73 and i10-index is 322 in Google Citation (Google Scholar). The total citation count of his papers raises above 18700+ times. Dr. Han is a Fellow of the UK Institution of Engineering and Technology (FIET). He has served on the Editorial Boards of up to 10 international journals, including the IEEE TII, IEEE TCCN, IEEE TVT, IEEE Systems, etc. He has guest-edited several special issues in IEEE Journals and Magazines, including the IEEE JSAC, IEEE Communications, IEEE Wireless Communications, Computer Networks, etc. Dr. Han has also served as chair of organizing and technical committees in many international conferences. He has been awarded 2020 IEEE Systems Journal Annual Best Paper Award and the 2017-2019 IEEE ACCESS Outstanding Associate Editor Award. He is a Fellow of IEEE.
\end{IEEEbiography}
\vspace{2ex}
\begin{IEEEbiography}[{\includegraphics[width=1in,height=1.25in,clip,keepaspectratio]{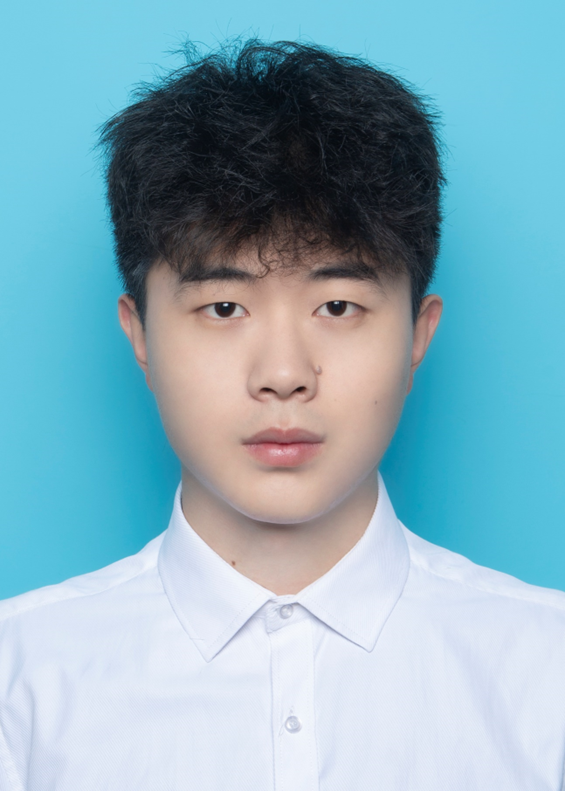}}]{Shengchao Zhu}
received his B.S. degree in Internet of Things Engineering from Hohai University,Changzhou, China, in 2023. He is currently pursuing the Ph.D. degree with the Department of Computer Science and Technology at Hohai University, Nan-
jing, China. His current research interests include swarm intelligence, swarm ocean, Multi-Agent Reinforcement Learning.
\end{IEEEbiography}
\vspace{2ex}
\begin{IEEEbiography}[{\includegraphics[width=1in,height=1.25in,clip,keepaspectratio]{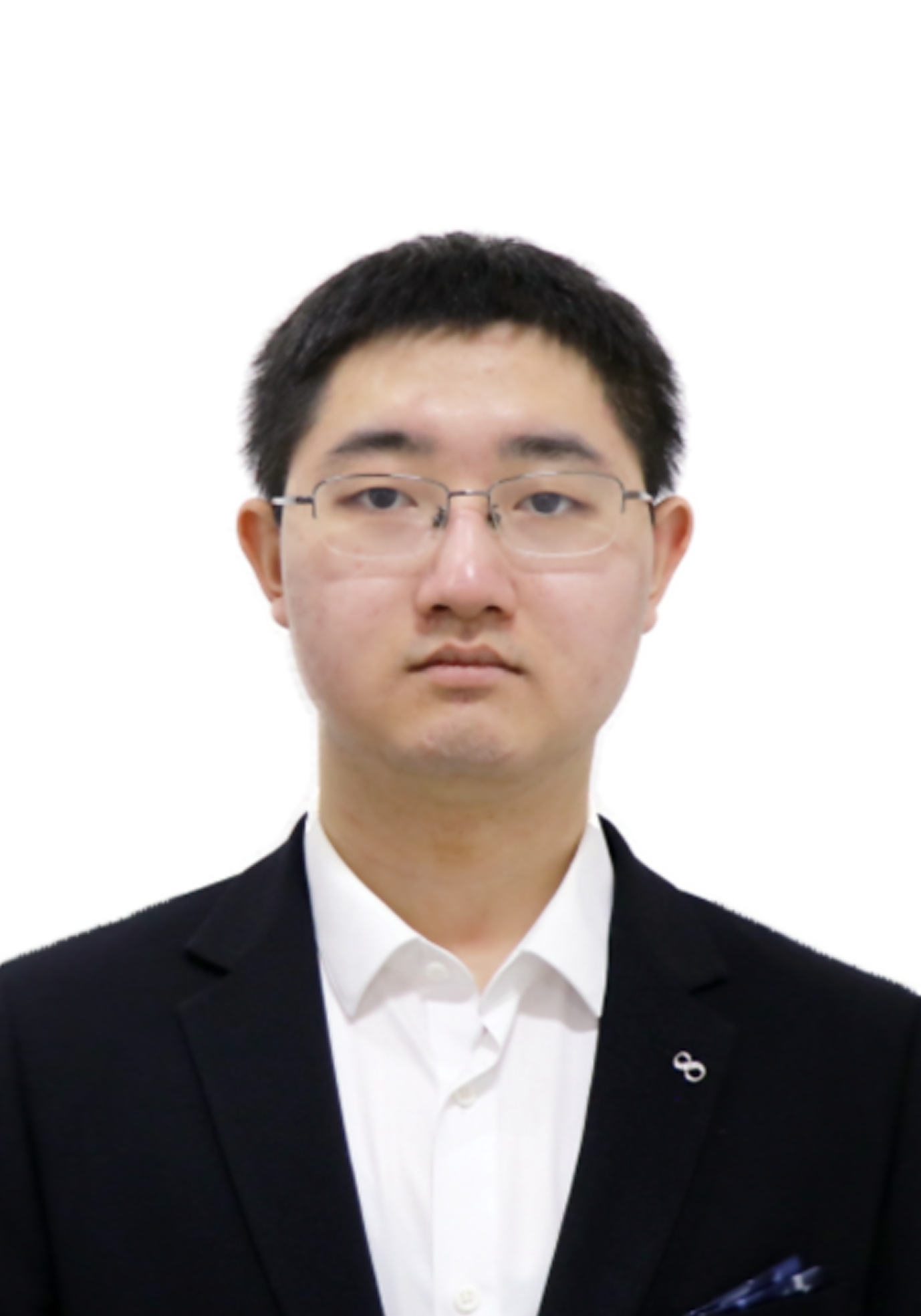}}]{Zhixian Li}
is currently pursuing a bachelor's degree at the Software College, Northeastern University, Shenyang, China. His research interests include natural language processing, computer vision and distributed computing.
\end{IEEEbiography}
\vspace{2ex}
\begin{IEEEbiography}[{\includegraphics[width=1in,height=1.25in,clip,keepaspectratio]{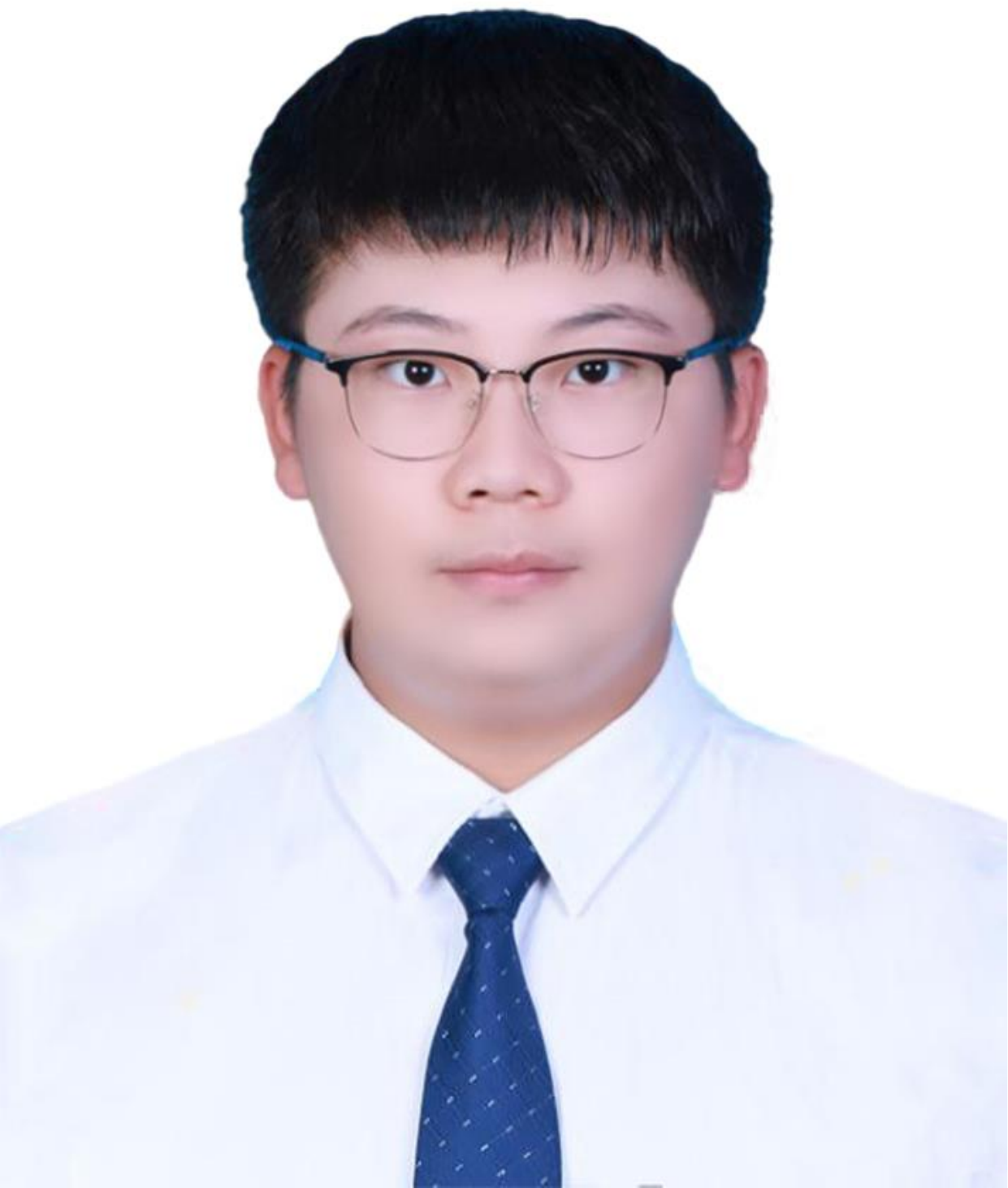}}]{Zhenyu Wang}
is currently pursuing a Bachelor's degree at the Software College, Northeastern University, Shenyang, China. His research interests include machine learning, software architecture and data mining.
\end{IEEEbiography}
\begin{IEEEbiography}[{\includegraphics[width=1in,height=1.25in,clip,keepaspectratio]{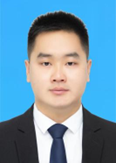}}]{Yunpeng Ma}
is currently an associate researcher with the College of Information Science and Engineering, Hohai University, Changzhou, China. He received his Ph.D. degree in Internet of Things Technology and Application from Hohai University, Nanjing, China, in 2019. In May 2020, he finished his work as a Postdoctoral Researcher with the Department of Computer Science and Technology, Hohai University, Nanjing, China. His current research interests include Intelligent visual perception, Biomimetic stereoscopic vision, 3D measurement and reconstruction. Dr. Ma has over 30 peer-reviewed journal and conference papers, in addition to 20 granted and pending patents.
\end{IEEEbiography}

\end{document}